\documentclass[fleqn,usenatbib]{mnras}
\usepackage{newtxtext,newtxmath}
\usepackage[T1]{fontenc}
\usepackage[normalem]{ulem}
\DeclareRobustCommand{\VAN}[3]{#2}
\let\VANthebibliography\thebibliography
\def\thebibliography{\DeclareRobustCommand{\VAN}[3]{##3}\VANthebibliography}

\usepackage{graphicx}	
\usepackage{amsmath}	
\usepackage{comment}
\usepackage{academicons}
\usepackage{xcolor}
\usepackage{url}

\newcommand{\re}[1]{\textcolor{black}{#1}}

\newcommand{\val}{\re{$(\mu_\alpha,\mu_\delta)= \left(-0.448\pm0.075, 0.058\pm0.078\right)$~mas~yr$^{-1}$}}
\newcommand{\valal}{\re{$(\mu_\alpha,\mu_\delta)= \left(-0.475\pm0.074, 0.008\pm0.078\right)$~mas~yr$^{-1}$}}

\title[HSC measurement of Sextans]{Study of structural parameters and systemic proper motion of Sextans dwarf spheroidal galaxy with Subaru Hyper Suprime-Cam data}

\author[A.~Tokiwa et al.]{
Akira~Tokiwa$^{1,2,3}$\thanks{E-mail: akira.tokiwa@ipmu.jp},
Masahiro~Takada$^{1,3}$, 
Tian~Qiu$^{1,2,3}$, 
Naoki~Yasuda$^{1}$, 
Yutaka~Komiyama$^{4}$, 
\ Masashi~Chiba$^{5}$,
\and
Kohei Hayashi$^{6,5,7}$
\\
$^{1}$Kavli Institute for the Physics and Mathematics of the Universe (WPI), 5-1-5 Kashiwanoha, Kashiwa-shi, Chiba, 277-8583, Japan\\
$^{2}$Department of Physics, The University of Tokyo, 7-3-1 Hongo, Bunkyo-ku, Tokyo 113-0033 Japan\\
$^{3}$Center for Data-Driven Discovery, Kavli IPMU (WPI), UTIAS, The University of Tokyo, Kashiwa, Chiba 277-8583, Japan \\
$^{4}$Department of Advanced Sciences, Faculty of Science and Engineering, Hosei University, 3-7-2 Kajino-cho, Koganei-shi, Tokyo 184-8584, Japan\\
$^{5}$Astronomical Institute, Tohoku University, Aoba-ku, Sendai 980-8578, Japan\\
$^{6}$National Institute of Technology, Sendai College, 48 Nodayama, Medeshima-Shiote, Natori, Miyagi 981-1239, Japan \\
$^{7}$Institute for Cosmic Ray Research, The University of Tokyo, Chiba 277-8582, Japan
}

\date{Accepted XXX. Received YYY; in original form ZZZ}

\pubyear{2023}

\begin{document}
\label{firstpage}
\pagerange{\pageref{firstpage}--\pageref{lastpage}}
\maketitle

\begin{abstract}
We use the Subaru Hyper Suprime-Cam (HSC) data to study structural parameters and systemic proper motion of 
the Sextans dwarf spheroidal galaxy at the heliocentric distance of 86~kpc, 
which is one of the most important targets for studies of dark matter nature and galaxy formation physics. 
Thanks to the superb image quality and wide area coverage, 
the HSC data enables a secure selection of member star candidates based on the colour-magnitude cut, 
yielding about 10,000 member candidates at magnitudes down to  $i\sim 24$. 
We use a likelihood analysis of the two-dimensional distribution of stars to estimate the structural parameters of Sextans taking into account the contamination of foreground halo stars in the MW, 
and find that the member star distribution is well-fitted by an elliptical King profile with ellipticity 
$\epsilon \simeq 0.25$ 
and the core and tidal radii of 
$R_c=(368.4\pm 8.5)$~pc and 
$R_t=(2.54\pm 0.046)$~kpc, respectively. 
Then using the two HSC datasets of 2.66 years time baseline on average, 
we find the systemic proper motions of Sextans to be 
\val 
which is consistent with some of the previous works using the {\it Gaia} data of relatively bright member stars in Sextans. 
Thus, our results give a demonstration that a ground-based, large-aperture telescope data 
which covers a wide solid angle of the sky and has a long time baseline, 
such as the upcoming LSST data, can be used to study systemic proper motions of dwarf galaxies. 
\end{abstract}



\section{Introduction}
\label{sec:introduction}
Dwarf satellite galaxies offer invaluable avenues for the exploration of fundamental processes involved in galactic formation and the historical assembly of the Milky Way's (MW) halo\citep{1999ApJ...522...82K,2017ARA&A..55..343B}.
This allows us to rigorously evaluate key model predictions of the $\Lambda$CDM structure formation paradigm on smaller scales. For example, it provides the basis for investigating a hypothesised universal dark matter density profile \citep{1994Natur.370..629M,1997ApJ...490..493N,1997ApJ...477L...9F},
and offers a route to probe the particle nature of dark matter through the pursuit of annihilation and decay signals\citep{1990Natur.346...39L,2013pss5.book.1039W,2015ApJ...801...74G,2016MNRAS.461.2914H}.


Sextans dwarf spheroidal galaxy (hereafter merely Sextans), ranks among the most luminous "classical" dwarf galaxy satellites of the MW, situated at a heliocentric distance of 86~kpc\citep{1990MNRAS.244P..16I,1991AJ....101..892M,1995MNRAS.277.1354I,1995AJ....110.2166M,2003AJ....126.2840L,10.1093/mnras/stx086}.
This galaxy is notable for its extensive spatial extent \citep{1995MNRAS.277.1354I,10.1093/mnras/stw949,10.1093/mnras/stx086,2018ApJ...860...66M}, which could potentially indicate tidal disruption, although this interpretation is complicated by the lack of clear evidence such as discernible tidal tails or S-shaped contours in Sextans.
From studies on the internal kinematics, Sextans is found to be a dark matter dominated system; the total mass is about $10^8M_\odot$, with a very large mass-to-light ratio ($M/L$) in the range of $M/L\sim 500$--$ 900$ depending on the aperture radii\citep{2007ApJ...667L..53W,2009AJ....137.3100W,2008Natur.454.1096S,2011MNRAS.411.1013B}.
Thus Sextans is one of the most important targets for exploring the nature of dark matter \citep[e.g.][]{2020ApJ...904...45H}.

on the other hand, ascertaining the orbital motions of dwarf galaxies enables us to extrapolate the mass profile of the Milky Way (MW), and to approximate the orbital histories of the satellite galaxies\citep{2018AA...619A.103F,McConnachie_2020}. To derive the orbital data of each dwarf galaxy, it is necessary to secure both systemic radial and tangential velocities - these are extrapolated from the velocities of constituent stars - along with their spatial coordinates relative to the MW centre.
Among these phase variables, proper motions prove to be the most cost-intensive to measure accurately, particularly for faint stars at significant heliocentric distances, a characteristic indeed exemplified by Sextans at 86 ~kpc. The introduction of $Gaia$ \citep{2018A&A...616A...1G} has revolutionised this situation, making precise proper motion measurements readily accessible for brighter stars. For Sextans, a multitude of studies have utilised $Gaia$'s proper motion measurements of probable member stars on the red giant branch and the blue horizontal branch (and blue stragglers) to deduce the systemic proper motion of Sextans\citep{2018AA...616A..12G,2018AA...619A.103F,McConnachie_2020,Li_2021,10.1093/mnras/stab1568}. However, these bright stars only represent a minor segment of the member populations, for instance, when compared with main sequence stars. Thus, it remains crucial to ascertain the systemic proper motion of Sextans by examining a larger sample of its members, namely fainter stars, inclusive of those on the main sequence.

Thus, the aim of this paper is to scrutinise the structural parameters and systemic proper motion of Sextans, utilising data from the Subaru Hyper Suprime-Cam (HSC)\citep{2018PASJ...70S...1M,2018PASJ...70S...2K}. This data was gathered via multiple passbands across different epochs, with an average time baseline of approximately 2.66 years. Owing to its exceptional image quality, large aperture, and expansive field-of-view, the Subaru HSC data facilitates a clear delineation of stars from galaxies and a reliable selection of Sextans member star candidates through colour-magnitude cuts, down to faint magnitudes of $i\sim 24$. Consequently, we can utilise an unprecedentedly large number of the selected stars to accurately estimate the structural parameters. Despite being a ground-based dataset, we will demonstrate that the Subaru HSC data permits an estimation of the systemic proper motion of Sextans, using the method developed in \citet{2021MNRAS.501.5149Q}, specifically for HSC data. To accomplish this, we utilise stars adequately exterior to Sextans to estimate the number density and systemic proper motions of foreground MW halo stars, thereby minimising contamination from these stars in our measurements. Our research provides evidence that a ground-based dataset can be leveraged to study the proper motions of stars, globular clusters, and dwarf galaxies, given a longer time baseline and control of systematic errors are accessible. This is notably pertinent to the upcoming Rubin Observatory's LSST \footnote{\url{https://www.lsst.org}}.


The structure of this paper unfolds as follows. Section~\ref{sec:data} delineates the specifics of the Subaru HSC data utilised in this study. In Section~\ref{sec:dataprocess}, we detail the processes of data manipulation, including star/galaxy segregation, object matching across the two datasets, and the recalibration of astrometric solutions crucial for the measurement of proper motion. Section~\ref{sec:results} presents the core findings of this research, namely the estimates of the structural parameters and the systemic proper motions of Sextans. The paper concludes with Section~\ref{sec:conclusion}. The code utilised for the experiments in this study can be found in the following repository: \url{https://github.com/atokiwaipmu/HSCSextansPMMeasurement}

%
\section{Data}
\label{sec:data}
\begin{figure}
    \centering
    \includegraphics[width=0.95\hsize]{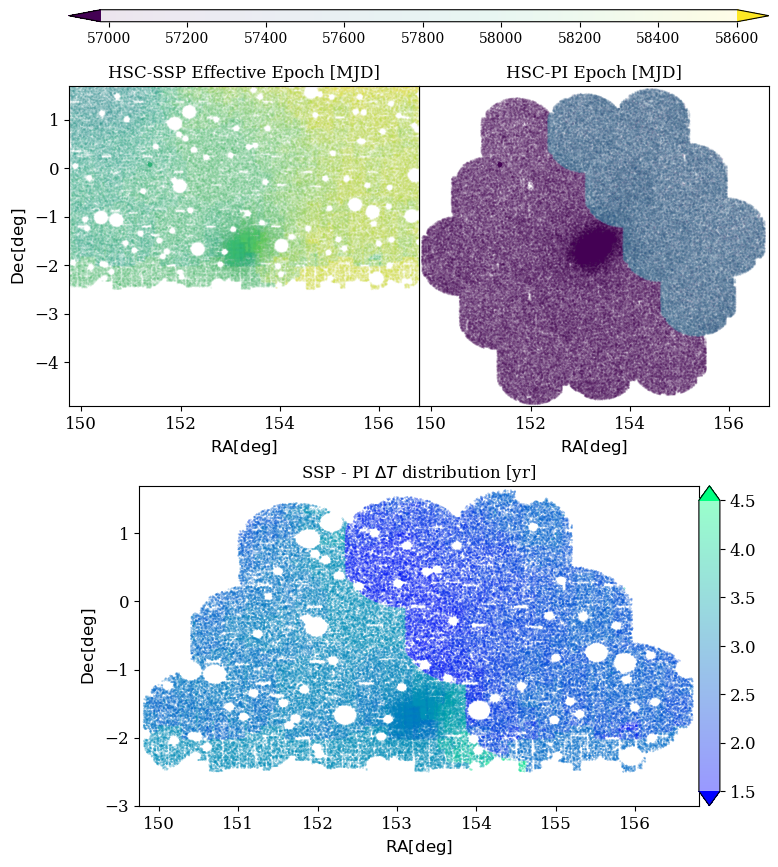}
    \caption{\re{
    Points depicted in the upper-left and upper-right panels represent the spatial distribution of stars for the {\it HSC-SSP} and {\it HSC-PI} catalogues, respectively, within the Sextans region. Regions encircled in white denote bright star masks (refer to the main text for details). The colour-coded points indicate the "effective" observed epoch (Modified Julian Date, or MJD) for each star, which are computed as the average of all exposures for each star. For the {\it HSC-PI} catalogue, the region coloured blue was captured in the i2-band at the 57481~MJD epoch, while the purple region was obtained from the i-band data at the 56980~MJD epoch. The bottom panel illustrates the spatial distribution of matched stars between the {\it HSC-SSP} and {\it HSC-PI} catalogues, with the colour code signifying the time difference between the observed epochs for each star. The average effective time baseline stands at approximately 2.66 years for the stars.}
    }
    \label{fig:mjdspatial}
\end{figure}

\begin{table}
    \begin{center}
  \begin{tabular}[t]{l|cc} \hline\hline
  Type & Number of objects \\\hline
  HSC-SSP \\
  \hspace{1em} all objects $(i_\mathrm{PSF} < 24)$ &1,444,206 \\ 
  \hspace{1em} stars used in this work  & 210,506 \\
  \hspace{1em} galaxies used in this work & 406,220 \\ \hline 
  HSC-PI\\
  \hspace{1em} all objects $(i_\mathrm{PSF} < 24.5)$ & 2,743,551 \\ 
  \hspace{1em} stars used in this work & 397,309 \\ 
  \hspace{1em} galaxies used in this work  & 879,791 \\ \hline 
  Matched stars for proper motion measurements &121,015 \\
  Matched galaxies used for calibration & 215,325 \\ \hline\hline
        \end{tabular}  
\end{center}
\caption{
This table presents the count of objects utilised in this study. 'All objects' represents the total count of detected objects with $i_{\rm PSF}<24$ or $24.5$ for the HSC-SSP and HSC-PI catalogues, respectively. A deeper magnitude cut for the HSC-PI data is applied owing to its shallower nature compared to the HSC-SSP data, ensuring we do not overlook any object present in the HSC-SSP catalogue due to greater photometric errors.  'Stars' and 'galaxies' within each catalogue are determined based on objects selected through the size threshold centred around $i_{\rm cModel}-i_{\rm PSF}$ and the colour-magnitude cuts (refer to the main text for more detailed information). The final rows signify the quantity of the matched stars utilised for the measurement of proper motion, and the count of matched galaxies employed for calibration purposes. 
}
\label{tab:numberofobjects}
\end{table}
This section delineates the specifics of the data employed in this research. To ascertain the proper motions of stars, we utilise imaging data captured at two distinct epochs. This enables the calculation of angular displacement for each star over the time interval, and subsequently, its proper motion. The entirety of the images employed were acquired through the Subaru Hyper Suprime-Cam (HSC), with details provided subsequently. The HSC camera, situated at the prime focus of the 8.2~m Subaru Telescope, is a wide-field imaging apparatus boasting a 1.77~deg$^2$ field-of-view \citep{2018PASJ...70S...1M,2018PASJ...70S...2K,2018PASJ...70S...3F,2018PASJ...70...66K}. Table~\ref{tab:numberofobjects} presents a summary of the number of objects incorporated in this research.  \re{Fig.~\ref{fig:mjdspatial} provides a depiction of the spatial distribution of 
"effective" observed epochs and time baseline for each star within the utilised HSC data.}
%
\subsection{HSC-SSP}
\label{sec:hsc-ssp}
\begin{figure}
    \centering
    \includegraphics[width=0.95\hsize]{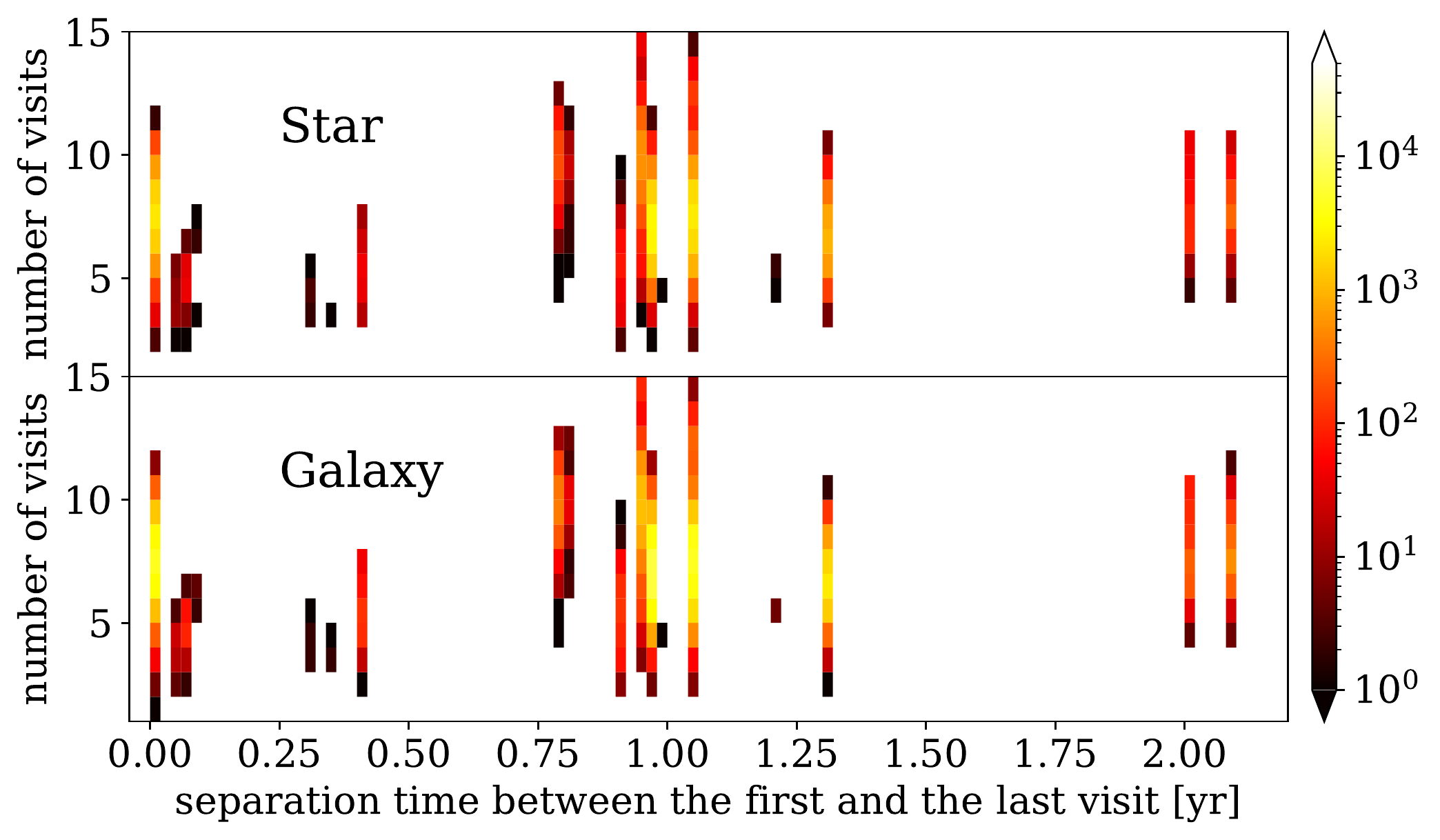}
    \caption{
    The HSC-SSP co-add images are compiled from data gathered over multiple visits to each field, with the co-addition conducted per HSC "tract" region. The colour assigned to each grid denotes the number of HSC tracts situated within a given value of the 2D space. This value is parameterised by the number of visits to each tract (represented on the $y$-axis) and the temporal separation between the first and last visits to the tract (represented on the $x$-axis). Employing this information regarding epochs and tracts, we calculate the mean epoch of the co-add image for each object within the HSC-SSP. This serves as a basis for measuring proper motion.
    }
    \label{fig:visitdelta}
\end{figure}
The initial HSC data utilised in our research is sourced from the HSC Subaru Strategic Program (HSC-SSP) survey, which commenced in 2014 and concluded early in 2022\citep{10.1093/pasj/psx066}. The HSC-SSP survey dedicated 330 nights on the Subaru telescope to execute a five-band ($grizy$) wide-area imaging survey spanning approximately $1,200$ deg$^2$. Each field received a total exposure time of 10~min for $gr$ and 20~min for $izy$, respectively. \re{The HSC-SSP data presents point source $5\sigma$ depths of $g\sim 26.5$, $r\sim 26$ and $i\sim 26$, respectively\citep{10.1093/pasj/psab122}.} Of the five filters, the $i$-band images were captured under favourable weather and seeing conditions (typically $0\farcs6$ seeing Full-Width Half-Maximum) \citep{10.1093/pasj/psx066,10.1093/pasj/psx130}. We utilize the catalogue from an approximate $20$ deg$^2$ field centred around Sextans, which is extracted from the internal S21A data release product based on the data gathered between March 2014 and January 2021. \re{This is accomplished by minimising the contamination of the star sample by galaxies (discussed further below), and also due to the fact that fainter stars with $i>24$ offer limited utility for the measurement of star proper motions, given the larger centroid determination errors associated with them as we will elucidate later.}

The HSC-SSP catalogue comprises primary photometric sources in the $g,r,i$ filters with $i_\mathrm{PSF} < 24$, amounting to approximately $1.78$ million objects. The sample selection query can be found in Appendix~\ref{ap:sspsql}. We utilise co-add images, which are assembled from varying exposures captured at different epochs \citep{2018PASJ...70S...5B}. All single exposures overlapping with a particular "tract" are projected onto an output plane centred on the tract. \re{Each tract represents a rectangular region, roughly 1\fdg68 on each side (slightly larger than the HSC field of view), with different tracts sharing an overlap of at least 1\arcmin on each side.} Then, the co-add images are generated by calculating the weighted average of the tract\citep{2018PASJ...70S...5B}. To establish the effective observed epoch for each co-add source, we compute its mean modified Julian date (MJD), based on the epochs of individual exposures (obviously, this is important for the HSC-SSP). Fig.~\ref{fig:visitdelta} displays the distribution of HSC-SSP co-add images within a 2D space, parameterised by the number of visits to each tract and the temporal interval between the first and last visits. The definition of star/galaxy will be elaborated upon in Sec.~\ref{sec:dataprocess}. The HSC-SSP co-add images, consisting of 1 to 16 exposures, are predominantly captured within a few days to a year. Details of the analysis pipeline utilised for data reduction can be found in \cite{10.1093/pasj/psx080}. The calibration of the photometry and astrometry is done against the Pan-STARRS1 \citep{2012ApJ...756..158S, 2012ApJ...750...99T,2013ApJS..205...20M, 10.1093/pasj/psx080, 2020ApJS..251....3M, 2020ApJS..251....4W, 2020ApJS..251....5M, 2020ApJS..251....6M,2020ApJS..251....7F}. \re{The astrometric calibration errors of stars documented in the catalogue exhibit superior precision, under $1$~mas for bright stars with $i < 20$, but decrease to approximately $10$~mas for fainter stars with $i\sim 24$ \citep[also see][]{10.1093/pasj/psab122}. Nevertheless, the astrometric calibration aligned with the Pan-STARRS1 catalogues doesn't account for stellar proper motions, thus it is crucial to execute recalibration of the astrometric solutions via the positions of matched galaxies across both catalogues (refer to Sec. \ref{sec:recalibration}).}

The process begins with individual detection in each band. Subsequently, footprints and peaks of identified sources across various bands are amalgamated to ensure consistency. This source measurement algorithm delivers independent measurements of positions and source parameters in each band for these peaks. A reference band is determined for each object, with the decision based on the signal-to-noise ratio and an aim to optimise the number of objects in the reference band. Lastly, the source measurement is re-executed with fixed position and shape parameters, which are grounded in the measurements from the reference band \citep{2018PASJ...70S...5B}.


We employ the "bright" star masks to exclude objects influenced by artefacts surrounding luminous stars, which can manifest as halo, ghost, and blooming effects \citep[refer to Section~4 of ][]{10.1093/pasj/psab122}. These masks are generated utilising the second data release from the $Gaia$ \citep{Li_2021}\citep[also see][for discussion on the bright star masks]{2022PASJ...74..421L}.

%
\subsection{The open-use HSC data of Sextans: HSC-PI}
The second HSC dataset we utilise comes from a PI-based HSC survey of the Sextans field, made accessible through both the Subaru open-use program and the Subaru/Keck Time-Exchange program, retrievable from the Subaru archive system \citep[SMOKA][]{2002ASPC..281..298B}. The data collection occurred on 2014-11-19, 2014-11-20, and 2016-04-03, with $g, i, i2$-band filters. Each field typically experienced an exposure time of approximately 210, 190, and 190 seconds for $g, i$, and $i2$ filters, respectively. This is slightly shallower than the HSC-SSP data, \re{by approximately $\sim 1$~mag}. The average seeing for HSC-PI data is around $0.8$--$1.2\arcsec$ for $g$-band, subject to variations in observational conditions, and about $\sim 0.7\arcsec$ for the $i$ and $i2$-band data. We undertook a re-analysis of the HSC-PI data, utilising the same version (v8.5.3) of the {\tt hscPipe} pipeline \citep{2010SPIE.7740E..15A} employed for the HSC-SSP data. This re-analysis yields approximately 2.74 million objects with $i_{\rm PSF}<24.5$ derived from the HSC-PI data. In order to determine the positions and observation time epochs of the objects, we rely on the centroid positions and the observed Modified Julian Date (MJD) in the $i$-band and $i2$-band. When an object is observed in both $i$ and $i2$-bands, preference is given to the data from the $i2$-band.
\section{Data Processing}
\label{sec:dataprocess}
\subsection{Star/galaxy separation}
\begin{figure}
    \centering
    \includegraphics[width=0.95\hsize]{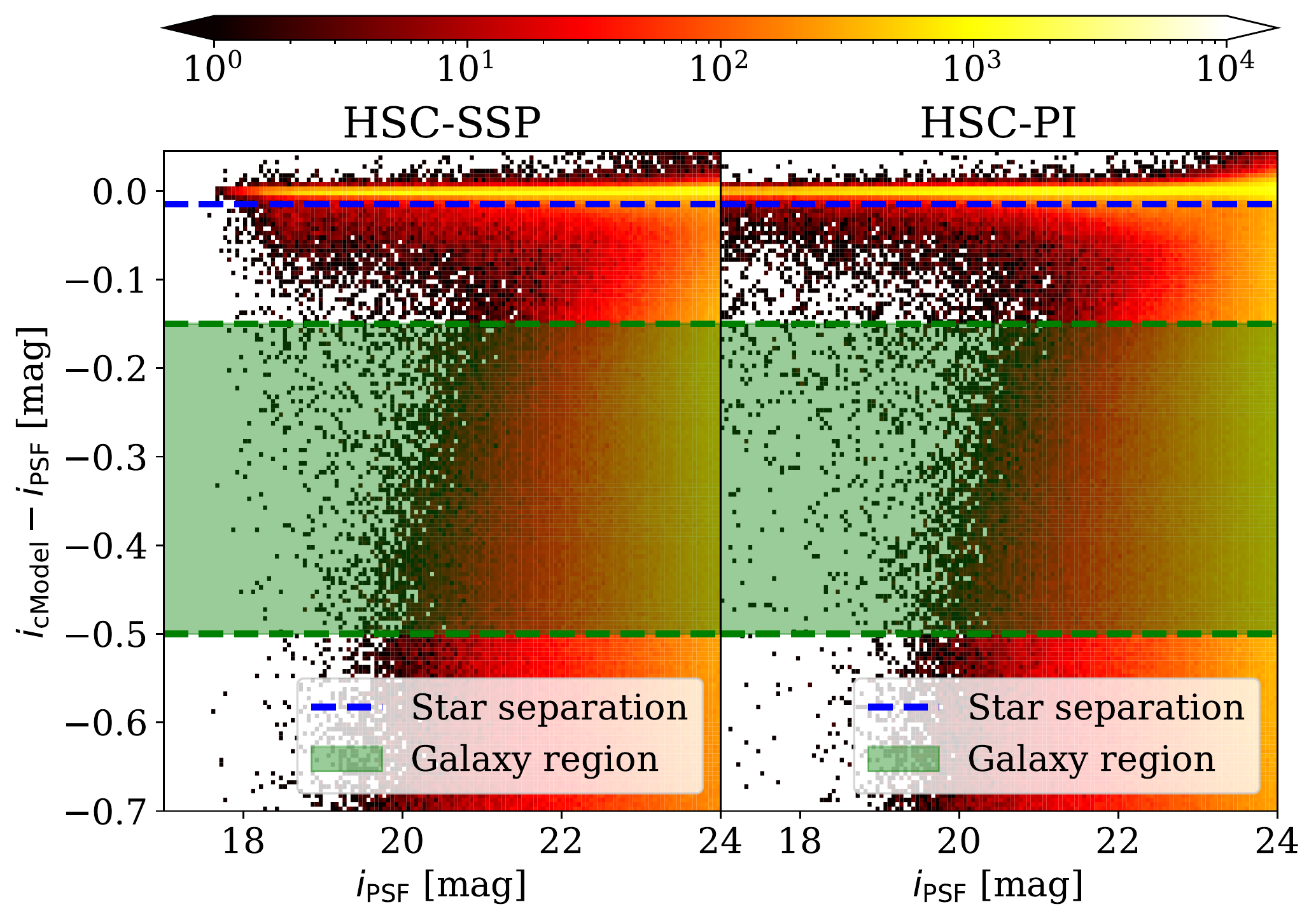}
    \caption{
    The distribution of all objects as a function of the $i$-band PSF magnitudes ($x$-axis) and the difference between the HSC $i$-band cModel and PSF magnitudes ($y$-axis). The colour corresponds to the number of objects within each grid of this two-dimensional space. We select stars satisfying the condition $(i_\mathrm{cModel} - i_\mathrm{PSF})> -0.015$, as indicated by the blue dashed line. To distinguish relatively compact galaxies from point sources, including stars, we apply a cut of $ -0.5 < (i_\mathrm{cModel} - i_\mathrm{PSF})< -0.15$, which is denoted by the green shaded region. These galaxies are utilised for the recalibration of the astrometric solutions for stars.
    }
    \label{fig:galstarsep}
\end{figure}
For an accurate measurement of the proper motions in Sextans, it is essential to first establish a reliable sample of stars. In pursuit of this, we utilise the high-resolution $i$-band data of HSC-SSP and distinguish point sources based on the PSF to cModel flux ratio \citep[see][for instance]{2004AJ....128..502A}. Fig.~\ref{fig:galstarsep} demonstrates the distribution of objects in the master catalogue, displayed in the plane defined by $i_{\rm PSF}$ versus $i_{\rm CModel}-i_{\rm PSF}$, where $i_{\rm PSF}$ and $i_{\rm CModel}$ respectively signify the PSF and cModel magnitudes. In this study, we establish the criterion $i_\mathrm{cModel} - i_\mathrm{PSF} > -0.015$, which aligns with the threshold set by the HSC pipeline for identifying point sources or potential stars. \re{Our confidence is high that the stars are securely selected using the aforementioned criterion. It is evident from the comprehensive study by \citet{2022PASJ...74..421L} that there is likely to be negligible galaxy contamination for stars with $i_{\rm PSF}<22.5$. Their research, which developed a galaxy shape catalogue for weak lensing studies, demonstrated that stringent testing of PSFs is crucial (as PSF characterization becomes degraded if galaxy contamination is present within the star catalogue). For star candidates where $22.5<i_{\rm PSF}<24$, residual contamination from galaxies may persist within the star catalogue. To address this, further cuts will be adopted to select Sextans member star candidates from the colour-magnitude diagram. We will discuss later that the contamination from galaxies is likely to be minimal.}

In addition to the star catalogue, we utilise a galaxy catalogue to rectify systematic errors within astrometric solutions for stars, juxtaposing the HSC-SSP and HSC-PI catalogues. We define galaxies as objects satisfying the condition $ -0.5 < i_\mathrm{cModel} - i_\mathrm{PSF} < -0.15$. The upper limit for the PSF-to-cModel flux ratio is set at $-0.15$, which significantly exceeds the star separation threshold of $-0.015$. We are confident that this selection criterion, even considering potential photometric errors, ensures a robust identification of galaxies, given the deep HSC photometry. This implies any migration of stars into this galaxy selection threshold owing to photometric errors would be negligible. The lower limit of $-0.5$ is imposed as galaxies with extensive spread are not beneficial for astrometry calibration due to their comparatively lower accuracy in centroid determination.
\subsection{Matching objects in the HSC-SSP and HSC-PI catalogues}
Pursuing the methodology delineated in \citet{2021MNRAS.501.5149Q}, the proper motion of each star is assessed via its angular offset between the HSC-SSP and HSC-PI catalogues over the given time baseline. This is carried out with respect to a reference frame established by the positions of galaxies. For the execution of this method, a matched star/galaxy catalogue is compiled between the HSC-SSP and HSC-PI catalogues.

We commence the matching process with each star or galaxy enlisted in the HSC-SSP catalogue, noted for its superior resolution and depth compared to the HSC-PI catalogue. Following this, a search is conducted for a corresponding object within the HSC-PI catalogue inside a radius of $1 \arcsec$. The matching radius of $1 \arcsec$ is deemed sufficient, considering that most of the stars under our scrutiny (particularly those in the Sextans, located 86kpc away) should exhibit substantially smaller angular offsets over our time baseline of approximately 4 years\footnote{For comparison, if a star at 1~kpc distance has a velocity greater than 2300~km~s$^{-1}$ with respect to us, the star has an angular offset larger than $1 \arcsec$ over 2~years. Most of the stars we are interested in are in greater distances than 1~kpc.}. For executing this matching process, we deployed the software tool {\tt TOPCAT}\footnote{\url{http://www.star.bris.ac.uk/~mbt/topcat/}}.

\begin{figure*}
    \centering
    \includegraphics[width=0.95\hsize]{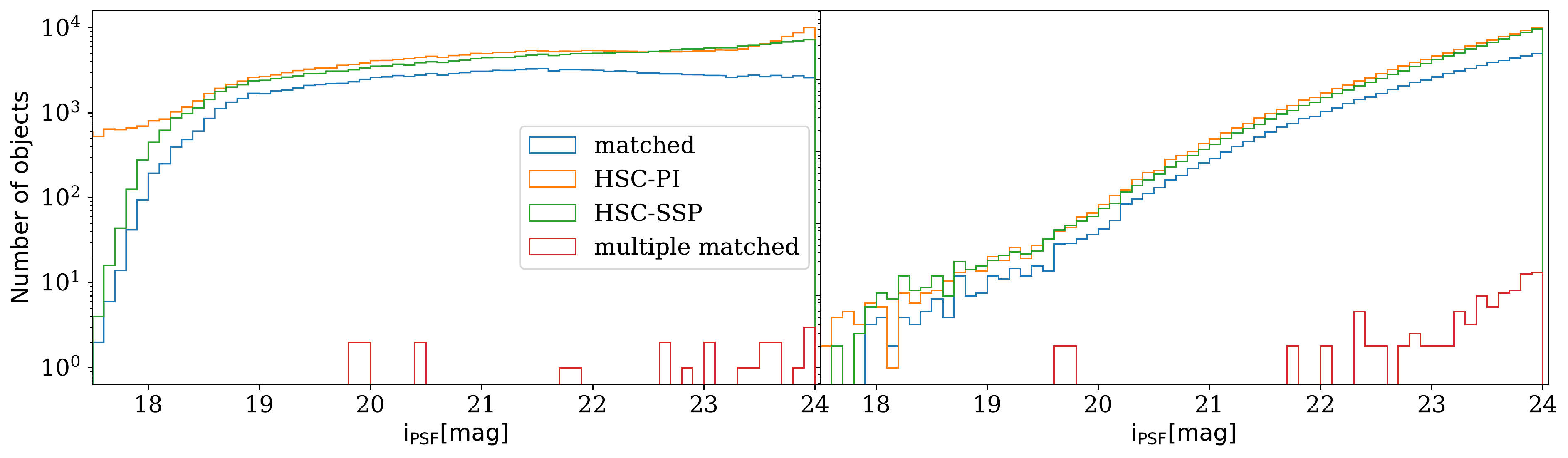}
    \caption{
    The histograms illustrating the magnitude distribution of the star/galaxy catalogues are presented in this figure. The left panel pertains to stars, while the right panel depicts galaxies. The HSC-PI catalogue objects are represented by orange histograms, and the HSC-SSP catalogue objects are marked by green histograms. Blue histograms display the distribution pertaining to matched catalogues. Additionally, the distribution of objects from the HSC-SSP that align with multiple objects within the HSC-PI is captured by the red histogram.
    }
    \label{fig:galstarmaghist}
\end{figure*}
Fig.~\ref{fig:galstarmaghist} presents the count of star and galaxy objects, categorised by their respective $i$-band PSF magnitudes for both the HSC-SSP and HSC-PI catalogues. To enhance the integrity of the matched entities, we exclude those manifesting considerable colour disparities between the HSC-SSP and HSC-PI catalogues. More specifically, we segregate star and galaxy objects based on their $i$-band PSF magnitudes and apply a $3\sigma$ clipping post-fitting a Gaussian distribution to the colour distribution within each bin. We employ the disparity between $(g-i){\rm SSP}$ and $(g-i){\rm PI}$ for this clipping procedure. This leads to the discarding of $9,083$ stars (making up $7\%$ of all stars) and $9,135$ galaxies (amounting to $4\%$ of all galaxies).

\re{We found that the HSC-PI catalogue's star objects are contaminated by fainter galaxies, which are efficiently eliminated by the superior-seeing HSC-SSP catalogue. This is evident in the rise in the count of stars for $i\gtrsim 23$ in the HSC-PI catalogue, attributed to galaxy contamination, which disappears post-matching. Additionally, the colour-magnitude diagram will be utilised to further mitigate galaxy contamination, as we will elaborate below.} A minor contamination is evident from objects with multiple matches, but their count is negligible compared to the single-matched entities. We will disregard these multiple-matched objects in the subsequent analysis. In sum, we have 121,015 matched stars and 215,325 matched galaxies, respectively.

%
\subsection{Recalibration}
\label{sec:recalibration}
Given that quasars and galaxies are remote entities, they should ostensibly exhibit no proper motions. Therefore, any discernible angular offsets in galaxy/quasar positions between the HSC-SSP and HSC-PI catalogues may be indicative of residual systematic errors in the astrometric solutions for both catalogues. To counter this, we employ the observed angular offsets of galaxies to recalibrate these systematic errors \citep{2013ApJ...766...79K,2021MNRAS.501.5149Q}. Following the strategy delineated in \citet{2021MNRAS.501.5149Q}, we construct a "recalibration" map. The size of the HSC CCD chip is approximately $0.196\times 0.096$~deg$^2$ ($4,176\times 2,048$~pixels, with an average pixel size of $0.168\arcsec$) \citep{2018PASJ...70S...1M}. Consequently, we divide the sky footprint into small grids, each measuring $0.05\times0.05$~deg$^2$. This grid size is chosen to be smaller than the size of the CCD chip and thus enables us to detect spatial variations in the systematic errors within the CCD chip. Additionally, it's noteworthy that each grid houses an ample number of galaxies for recalibration (on average, each grid houses roughly 25 galaxies). In every grid, we calculate the average angular offset of the matched galaxies between the HSC-SSP and HSC-PI, following a 3$\sigma$ clipping based on their angular offset distribution within that grid. We then divide the angular offsets by the average time separation, calculated according to the methodology outlined in Section~\ref{sec:data}, to translate the angular offset into units of proper motions in R.A. and Dec. directions, denoted in mas\,yr$^{-1}$, for each grid. \re{These errors are computed from the 25th, 50th, and 75th percentiles of the angular offset distribution within the grid, according to Eq.~(1) in \citet{2021MNRAS.501.5149Q}}.
\subsection{Proper motion measurement of stars}
\label{sec:pmmeasure}
\begin{figure*}
    \centering
    \includegraphics[width=0.95\hsize]{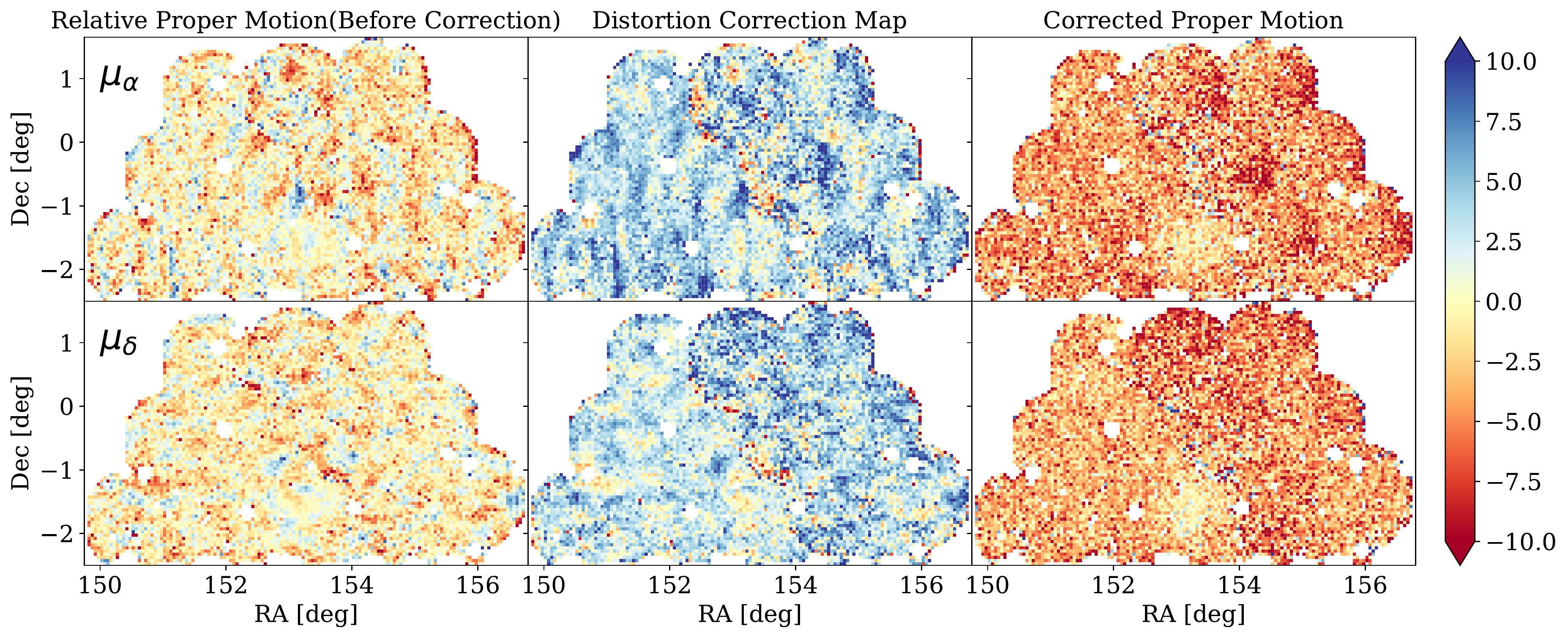}
    \caption{
    {\it Left panel}: The mean "net" proper motion of stars within each grid is calculated by averaging the angular offsets of star positions within each grid between the HSC-SSP and HSC-PI catalogues. The angular offsets are then converted into proper motion units, [mas\,yr$^{-1}$], by dividing by the time baseline calculated for each object, as depicted in Fig.~\ref{fig:visitdelta}. The maps are compiled from a total of 8720 non-zero grids, each measuring $0.05 \times 0.05 ~\mathrm{deg}^2$ in size. Bright star masks, shown as white circles, are not incorporated into subsequent analyses.
    {\it Middle panel}: This is the recalibration map of astrometric solutions between the HSC-SSP and HSC-PI data, estimated from the angular offsets of galaxies within each grid and expressed in [mas~yr$^{-1}$] units. On average, each grid contains approximately 25 galaxies.
    {\it Right panel}: The proper motions of stars within each grid, adjusted for the recalibration of astrometric solutions, are shown. This is achieved by subtracting the recalibration (shown in the middle panel) from the net motions (shown in the left panel).
    }
    \label{fig:pmmap}
\end{figure*}
For each pair of matched stars, we initially estimate their proper motion by first calculating the angular offset between their positions in the HSC-SSP and HSC-PI catalogues, and then dividing it by the time difference between the two observations. The time baseline used for the proper motion measurements is determined by the interval between the mean HSC-SSP and HSC-PI MJD dates for each matched pair. For the matched pairs of objects (both stars and galaxies), the time baseline varies from 1.66 to 4.12 years. The average time baseline spans approximately 2.66 years, as can be observed in Fig.~\ref{fig:mjdspatial}.

The initial estimation of proper motions for the matched stars is depicted in the left panel of Fig.~\ref{fig:pmmap}. This map reveals distinct spatial patterns mirroring the field-of-view (FoV) of HSC, a consequence of neither the HSC-SSP nor HSC-PI data having prior corrections for the proper motions of stars before co-addition. As such, residual systematic errors in astrometric solutions likely persist in both catalogues.

To mitigate the systematic errors in astrometric solutions, we employ the matched galaxy catalogues, capitalising on the inherent quality that galaxies do not exhibit any proper motions, thus serving as reference coordinates in both datasets. This essentially involves utilising the angular offset map introduced in Section~\ref{sec:recalibration}. The middle panel of Fig.~\ref{fig:pmmap} presents a map of the mean angular offsets for matched galaxies within each grid. A spatial structure bearing resemblances to the FoV of HSC pointings and depth variations, much akin to what is observed in the left panel, is evident.

To recalibrate the proper motion measurements for each matched star, we identify the nearest grid from the galaxy offset map, and subsequently subtract the mean annual angular offset of galaxies from the star's observed proper motion. This yields our fiducial dataset for proper motion measurements. The recalibrated proper motions are illustrated in the right panel of Fig.~\ref{fig:pmmap}, revealing an absence of discernible spatial structure. The proper motions, excluding those in the Sextans region, present a spatially homogeneous distribution, indicative of the proper motions of foreground halo stars in the Milky Way. This motion includes both the intrinsic proper motion and the reflex motion resulting from the Solar system's own motion.
%
\section{Results}
\label{sec:results}
In the following section, our initial step involves selecting Sextans member stars through a colour-magnitude diagram, followed by the estimation of Sextans' structural parameters utilising three distinct density profile models. Subsequently, we conduct a measurement of Sextans' systemic proper motion and compare them with the results of previous studies.
%
\subsection{Selection of Sextans member stars based on colour-magnitude diagram}
\label{sec:CMD}
To confidently measure the proper motion of the Sextans dwarf galaxy, we distinguish Sextans member star candidates from MW foreground stars utilising their colour-magnitude diagram (CMD) \citep[also see][for a similar methodology]{10.1093/mnras/stw949}. To identify member star candidates via CMD, we focus on the "core" region extending to a $0.5^{\circ}$ radius from Sextans' centre, roughly corresponding to the half-light radius as estimated in \cite{1995MNRAS.277.1354I}, approximately $27.6\arcmin$. Throughout this section, we utilise each star's magnitude that has been corrected for Galactic dust extinction.
%
\begin{figure}
    \centering
    \includegraphics[width=0.95\hsize]{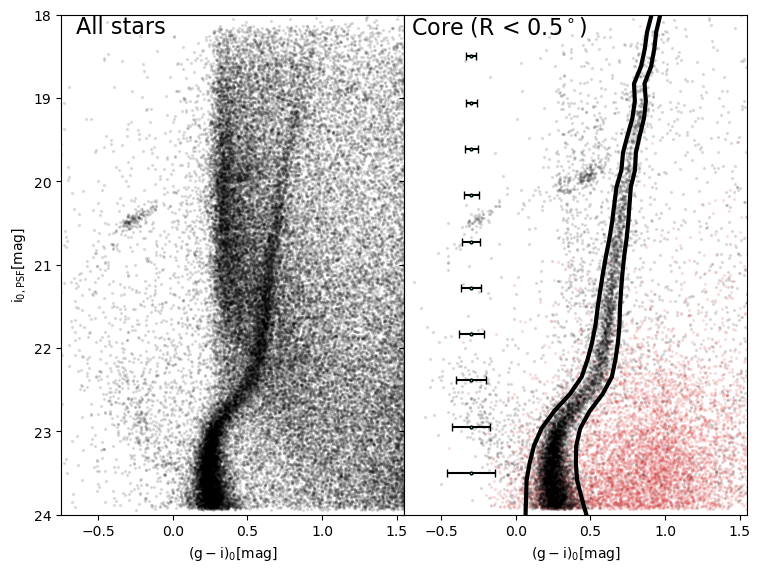}
    \caption{
    The colour-magnitude diagram (CMD) for stars in all data regions (left panel) and the core region (right) is depicted, where the core region is defined by a 0.5~degree radius from the Sextans centre. The core region comprises a total of 12,377 stars, while our selection of Sextans member candidates, demarcated by the two solid lines, comprises 7,951 stars. The points with error bars displayed represent the distribution width of stars in each magnitude. \re{The red points signify the CMD of galaxies within the core region used for calibration. Approximately 10\% of galaxies are found within the two bold lines.}
    }
    \label{fig:member}
\end{figure}
Fig.~\ref{fig:member} illustrates the $(g-i)$ colour and $i$-band magnitude diagram (CMD) for all stars within the Sextans' core region. For comparison purposes, the CMD for stars across all fields of our dataset is also presented. The CMD extends to the main-sequence (MS) turn-off, and exhibits a pronounced red giant branch (RGB) descending from $i \sim 18$ to $i \sim 22.5$, a blue and a red horizontal branch (BHB and RHB, respectively) situated at $i = 20 \sim 21$, a sub-giant branch (SGB), and blue stragglers (BSs) trailing down to $i\sim 23$. As discernible from the left panel, numerous stars align with the colour-magnitude sectors of the Sextans RGB, SGB, and MS stars, particularly stars with colour $g-i\sim 0.3$-$0.5$ that are likely faint MS stars in the MW foregrounds \citep[also refer][for a similar discussion]{10.1093/mnras/stx086}. Conversely, stars corresponding to RHB, BHB, and BSs are relatively scarce. Therefore, we delineate boundaries to selectively identify RGB, SGB, and MS star candidates of the Sextans. Firstly, we disregard horizontal branch stars, $i \sim 20$, and apply colour boundaries, $0 < \left(g-i\right) < 1$, to pinpoint stars within these boundaries for Sextans' core region.

\begin{figure}
    \centering
    \includegraphics[width=0.98\hsize]{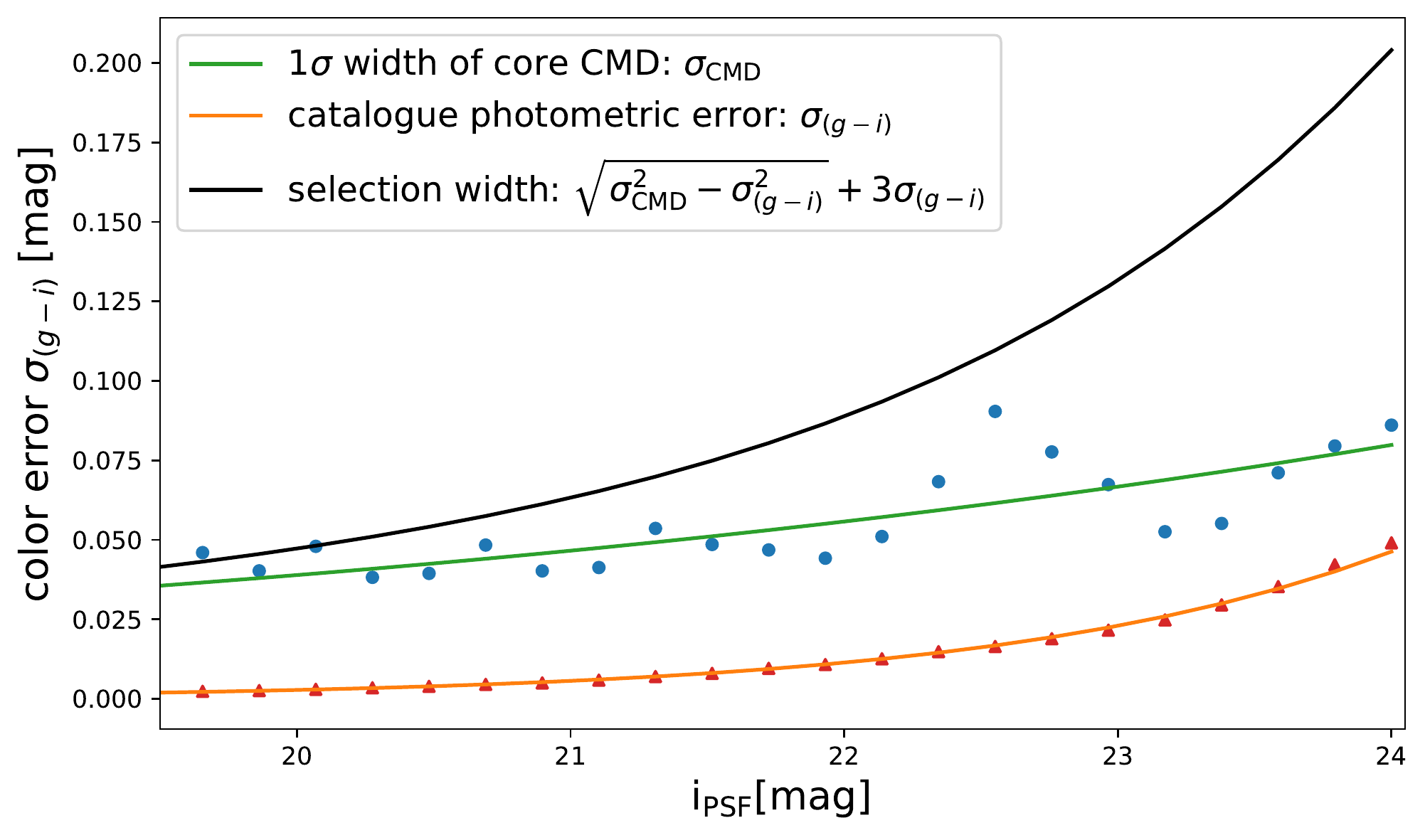}
    \caption{
    \re{The illustration of colour errors in relation to magnitude is depicted here. Red triangles alongside the accompanying orange line represent the catalogue photometric error and its corresponding fitted line. Blue circles, together with their corresponding green line, signify the 1-$\sigma$ width of the core CMD and its fitted line. The black line, used in this study, stands for the selection width.}
    }
    \label{fig:CMDwidth}
\end{figure}

Secondly, we compute the median colour $(g-i)$ for each magnitude bin, incrementing by $0.05$mag. \re{In an effort to define an accurate selection width, we fit the colour distribution for stars in each magnitude bin using a Gaussian function, centred at the median of the colour. The 1-$\sigma$ width is then fitted with an exponential function, symbolised as $\sigma_\mathrm{CMD}$. Similarly, we fit the catalogue photometric error, signified as $\sigma_{\left(g-i\right)}$, with an exponential function. The ultimate selection width is computed employing the formula $\sqrt{\sigma_\mathrm{CMD}^2 - \sigma_{\left(g-i\right)}^2} + 3\sigma_{\left(g-i\right)}$ (refer to Fig.\ref{fig:CMDwidth}).}
%

Thirdly, we designate stars that fall within the aforementioned selection width as probable members of the Sextans. Specifically, we perform a cubic interpolation of the median colour $(g-i)$ in each magnitude bin as a function of magnitude, subsequently selecting stars that fall within the selection width range (as denoted by the two bold solid lines in the right panel of Fig.~\ref{fig:member}). This outlines our method for selecting Sextans members.

\begin{figure}
    \centering
    \includegraphics[width=0.98\hsize]{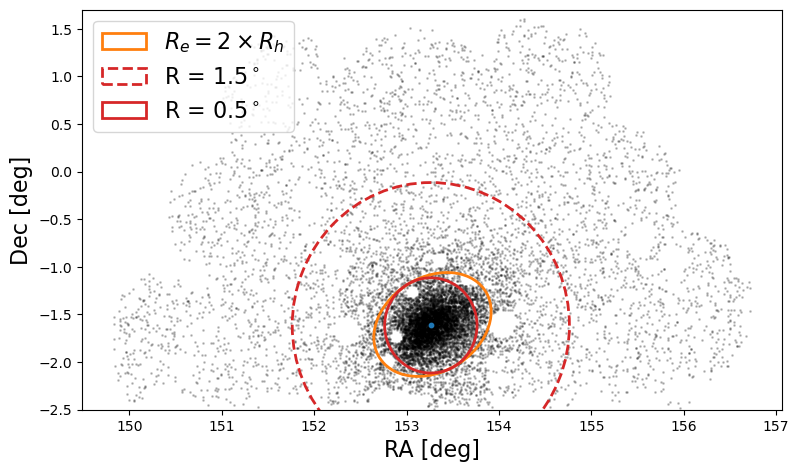}
    \caption{
    The spatial distribution of stars chosen based on the colour-magnitude criterion presented in the right panel of Fig.\ref{fig:member} is depicted in this figure. The concentration of stars indicative of Sextans member stars is unmistakable. The blue dot represents the centre of Sextans. The area inside the red solid circle, with a radius of 0.5~degrees from the centre, is where stars are selected to establish the colour-magnitude cut in Fig.\ref{fig:member}. The red dashed circle marks the radius within which stars are employed for estimating the structural parameters of Sextans. The orange solid ellipse corresponds to twice the half-light radius ($R_h$), where $R_h$ is derived from the fit to the Plummer profile (see text for further details). We will engage stars within $2R_h$ to ascertain the proper motion of Sextans.
    }
    \label{fig:numspa}
\end{figure}

\begin{figure}
    \centering
    \includegraphics[width=0.98\hsize]{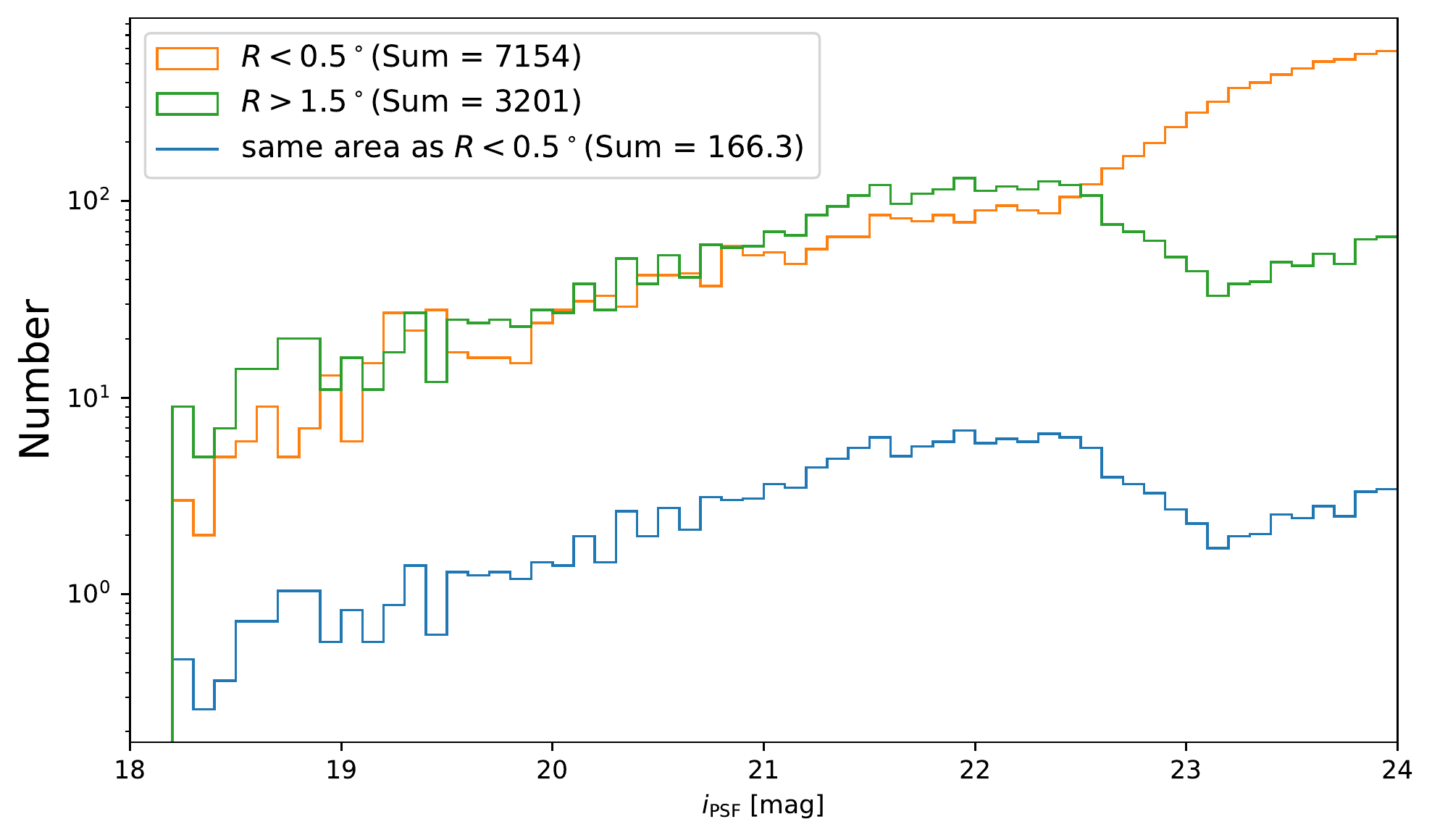}
    \caption{
    \re{The orange histogram portrays the magnitude distribution of 7,154 stars concentrated in the core region within a radius of $R < 0.5^\circ$ from the Sextans centre (as depicted in Fig.~\ref{fig:numspa}). The green histogram outlines the magnitude distribution of 3,201 stars situated beyond Sextans, at distances of $R > 1.5^\circ$. For the sake of comparison, a blue histogram is included, illustrating the magnitude distribution of stars in an area outside of Sextans roughly equivalent to the core region.}
    }
    \label{fig:maghist}
\end{figure}

By employing the aforementioned criteria to identify Sextans member candidates over the entire Sextans field, which includes areas outside the core region, we have successfully identified 13,792 stars. The spatial distribution of these selected stars is depicted in Fig.~\ref{fig:numspa}.
%
The figure evidently displays a substantial star concentration around Sextans, affirming the effectiveness of our member star selection process. Moreover, it's apparent that the star count beyond the Sextans region is significantly lower. \re{For instance, considering an area similar in size to the core region but situated outside Sextans, we identify only 166 stars, a stark contrast to the 7,154 stars found within the core region (refer to Fig.~\ref{fig:maghist})}. These residual stars in the outer region are presumably foreground halo stars from the MW, and their data will be employed to approximate contamination caused by these foreground MW halo stars in our measurements. Details on this process will be expounded upon in the subsequent sections.

%
Fig.\ref{fig:maghist} presents the magnitude distribution of stars in both the core region of Sextans and areas external to Sextans, following the CMD selection. It is evident that the core region is predominantly inhabited by faint stars with $i_{\rm PSF}>22.5$, characteristic of Sextans member candidates. Given Sextans' considerable heliocentric distance of approximately 86kpc (with a distance modulus of 19.7), member stars are expected to appear fainter than their foreground MW halo star counterparts. In contrast, regions outside of Sextans (separations exceeding 1.5~deg.) are characterised by the dominance of stars with $i\sim 22$, indicating an overall brighter population due to their closer distances.
%
\subsection{Structural properties}
\label{sec:structure1d}
Before proceeding to the proper motion measurement, we begin by estimating the structural parameters of Sextans. We adopt the maximum likelihood-based methodology presented in \citet{2008ApJ...684.1075M} and \citet{2018ApJ...860...66M} \citep[also see][for the original work]{1998AJ....115.2359K,2006AJ....131..375W}. The likelihood of deriving structural parameters, denoted as ${\bf p}$, from the observed angular positions of member stars, represented as ${\boldsymbol{\theta}_i}$, within the footprint is given by:
%
\begin{align}
\label{eq:totalllikelihood}
{\cal L}(\{\boldsymbol{\theta}_i\}|{\bf p})=\prod_{i}\ell_i(r_i |{\bf p})\, ,
\end{align}
Here, ${\bf p}$ represents a collection of parameters in the model profile for member stars (details to follow), $r_i$ stands for the angular separation of the $i$-th star from the centre of Sextans (further details will be given subsequently), $\ell_i$ signifies the likelihood for the $i$-th star, and the product $\prod_i$ encompasses all the stars incorporated in the fitting. The likelihood $\ell_i$ for each individual star, indexed by $i$, can be expressed as:
%
\begin{align}
    \mathscr{l}_i(r_i|{\bf p})&=
    \frac{N_{\rm tot}-S_{\rm fit}\Sigma_b}{A}{\Sigma}_s(r_i,{\bf p})+\Sigma_b,
    \label{eq:likelihood_i}
\end{align}
In the above expression, $N_{\rm tot}$ signifies the total count of stars involved in the fitting process, while $S_{\rm fit}$ denotes the effective area ([deg$^2$]) utilised in the fitting. ${\Sigma}_s(r_i,{\bf p})$ represents the model of the normalised radial profile at the position $r_i$ from the centre, and $\Sigma_b$ symbolises the average number density of foreground MW stars ([deg$^{-2}$]). It should be noted that we employ dimensionless units for ${\Sigma}_s(r_i,{\bf p})$. \re{Stars residing outside the tidal radius of Sextans, approximately $1.5$ degrees, are predominantly foreground MW stars, contributing minimally to the likelihood estimation of the structural characteristics.} \re{Including stars beyond the 1.5-degree has little impact on the parameters, barring the characteristic radius, which deviates by less than 1$\sigma$. The characteristic radius, expressed in terms of the half-light radius, varies by approximately 1\arcmin, or $3\sim 4 \sigma$, if all data is incorporated. Thus, we exclusively consider} stars located within a 1.5-degree radius from the centre of Sextans (refer to Fig.~\ref{fig:numspa}). After applying the colour-magnitude filter as per Fig.~\ref{fig:member}, a total of 10,591 stars remain within this region.
The likelihood $\mathscr{l}_i$ for the $i$-th star considers the probabilities of the star being a Sextans member as well as a foreground MW star. In terms of the model profile, we utilise the ellipsoid profile throughout this paper. In such instances, the normalisation constant $A$ is defined as follows:
%
\begin{align}
A\equiv (1-\epsilon) \int_0^{\infty}2\pi R\mathrm{d}R~ {\Sigma}_s(R,{\bf p}),
\end{align}
In the equation above, $\epsilon$ represents the ellipticity of the model profile. The constant $A$ carries the unit [deg$^2$] while $R$ symbolises the radius from the centre in elliptical coordinates, corresponding to the major and minor axes. The elliptical radius $r_i$ for the $i$-th star, calculated from its angular position, is determined as:
%
\begin{eqnarray}
r_i^2  &=&  \frac{\left(X_i\cos\theta \re{-} Y_i\sin\theta\right)^2}{\left(1- \epsilon\right)^2} + \left(X_i\sin\theta \re{+} Y_i\cos\theta\right)^2 \, 
\end{eqnarray}
Here, $X_i= (\alpha_i -\alpha_0)\cos\delta_0$ and $Y_i=\delta_i-\delta_0$, where $\alpha_0$ and $\delta_0$ are the  R.A. and Dec. of the presumed centre of Sextans, respectively. Additionally, $\theta$ represents the position angle of the major axis of the ellipsoid profile, \re{measured East from North in the sky coordinate system.}

For the purpose of this paper, we will be considering three distinct profiles for $\Sigma_s$ in Eq.~(\ref{eq:likelihood_i}): the Plummer profile \citep{1911MNRAS..71..460P}, the exponential profile, and the King profile \citep{1962AJ.....67..471K}.
%
\begin{align}
&\Sigma_{\mathrm{exp}}(R;R_E) = \exp{\left(-\frac{R}{R_E}\right)},\nonumber\\
&\Sigma_{\mathrm{Plummer}}(R;R_p) = \left(1+\frac{R^2}{R_p^2}\right)^{-2}, \nonumber\\
&\Sigma_{\mathrm{King}}(R;R_c,R_t) = \left[
\left(1+\frac{R^2}{R_c^2}\right)^{-1/2}
-\left(1+\frac{R_t^2}{R_c^2}\right)^{-1/2}
\right]^2,
\end{align}
In these equations, $R_E$ and $R_p$ stand for the scale radius of the exponential and Plummer profiles respectively, whereas $R_c$ and $R_t$ represent the core and tidal radii for the King profile. It is essential to note that the King profile is strictly defined for $R<R_t$. The exponential and Plummer profiles are both characterised by five model parameters: the coordinates for the centre (R.A. and Dec.), the ellipticity ($\epsilon$), the position angle of the major axis ($\theta$), and the scale radius ($R_E$ or $R_p$). The King model, on the other hand, is defined by six model parameters, with the first four parameters identical to the others, supplemented by $R_c$ and $R_t$. Notably, $R_p$ is equivalent to the half-light radius $R_h$ if the star distribution follows a Plummer profile, a parameter we will frequently refer to in this paper.

In this study, we employ stars residing within a $1.5$ degree radius from the reference centre, (R.A., Dec.)$=(153.26,-1.614)$, to derive the structural parameters through model fitting, following the approach set out in \cite{2016MNRAS.460.4492D}. To determine the effective area, $S_{\rm fit}$, we first segment the fitting region into grids, each spanning an area of $0.025\times 0.025~$deg$^2$. We then aggregate all the grids that encompass galaxies to approximate $S_{\rm fit}$. This method enables the efficient exclusion of grids falling within bright star masks. We employ galaxies to distinguish the surviving grids due to the abundance of galaxies compared to stars, post the magnitude and colour cut selection of Sextans members (refer to Sec.~\ref{sec:CMD} and the right panel of Fig.~\ref{fig:member}). It's noteworthy that we opted for a smaller-sized grid of $0.025\times 0.025$deg$^2$ as opposed to the $0.05\times 0.05$deg$^2$ grids employed for the astrometry calibration (Sec.~\ref{sec:recalibration}), as the reduced area of the grids enables a more precise estimation of the effective area. Concurrently, we determine the foreground number density, $\Sigma_b$, using the total count of stars located outside the fitting region, yielding $\Sigma_b\simeq 218.41\mathrm{deg^{-2}}$. \re{Despite the notion that the mean foreground number density should ideally be incorporated within the model fitting, we determine it independently due to its correlation with the profile's scale radius.}

To infer the structural parameters, we utilise the Markov chain Monte Carlo (MCMC) method facilitated by the openly available {\tt emcee} software \citep{Foreman-Mackey_2013}. We set the number of walkers and the maximum time step at $40$ and $10^4$, respectively. We choose to use a non-informative flat prior for each parameter as detailed in Table~\ref{tab:structural1d}, defining "non-informative" as the condition where the marginalised $1\sigma$ error of each parameter is considerably smaller than the flat prior range. The inferred parameters are compiled in Table~\ref{tab:structural1d} and a comprehensive parameter space corner plot is presented in Appendix~\ref{ap:wedge}. Moreover, the best-fit Plummer and King models are displayed in the upper panel of Fig.~\ref{fig:radpro}, juxtaposed with the 1D radial profile depicting the projected star number density as a function of the radii from the centre. For this purpose, the best-fit model derived from a 2D likelihood analysis is used to compute the 1D radial profile in relation to the radius from the centre in the optimal elliptical coordinate. As indicated in the figure, the best-fit model closely approximates the observed radial profile.
Our result can be compared to the previous works: 
$(R_p, R_c,R_t)=(16.5\arcmin\pm0.10\arcmin, 20.1\arcmin\pm 0.5\arcmin,60.5\arcmin\pm 0.6\arcmin)$ \citep{2018ApJ...860...66M};
$(R_p, R_c,R_t)=(26.60\arcmin\pm0.43, 17.91\arcmin\pm0.65\arcmin, 120.5\arcmin\pm7.7\arcmin)$ \citep{10.1093/mnras/stx086};
$(R_p, R_c,R_t)=(23.0\arcmin\pm0.4\arcmin, 26.8\arcmin\pm1.2\arcmin, 83.2\arcmin\pm7.1\arcmin)$ \citep{10.1093/mnras/stw949};
\re{$(R_p, R_c,R_t)=(22.8\arcmin\pm0.7\arcmin, 13.8\arcmin\pm0.9\arcmin, 120\arcmin\pm20\arcmin)$ \citep{2018A&A...609A..53C}}. 
Our findings are largely in alignment with those presented by \citet{2018A&A...609A..53C}. The method we employ carries several advantages, notably our inclusion of fainter stars with $i_{\rm PSF}\simeq 24$, and our study of a substantially wider region surrounding Sextans. These elements enable us to more effectively estimate the foreground MW star contamination. Previous studies have relied upon brighter stars or data confined to a smaller area around Sextans. The spatial distribution of Sextans members, as calculated through our methodology, will be subsequently utilised in the estimation of Sextans' systemic proper motion. Through a comparison of the total likelihood, denoted as ${\cal L}$, for the best-fit model across the three profiles, we observed that the King model returns a marginally higher value. Consequently, we will employ the best-fit King profile in determining the 1D profile for Sextans' proper motion measurement, as detailed in Section~\ref{sec:pm_measurements}.
%
\begin{table*}
    \centering
    \begin{tabular}{llllll} \hline\hline
    Model&RA$_c$[deg] &DEC$c$[deg] & $\theta$[deg] & ellipticity: $\epsilon$ & $R_{\mathrm{scale}}$  \\ \hline
    Prior & $[149.75,156.80]$ & $[-2.50,1.70]$ & $[0, 90]$ & $[0, 1]$ & $R_{p, E, c}: [0\arcmin, 50\arcmin], R_t: [0\arcmin, 150\arcmin]$\\ \hline
    Plummer&$153.280\pm0.003$&$-1.605\pm0.003$ & $60.681\pm1.310$ & $0.257\pm0.011$ & $R_p: 20\farcm145\pm0\farcm232 = R_\mathrm{HL}$  \\
    Exponential&$153.280\pm0.003$&$-1.603\pm0.003$ & $61.471\pm1.342$& $0.250\pm 0.011$ & $R_E: 12\farcm347\pm0\farcm137, R_\mathrm{HL}: 20\farcm736\pm0\farcm234$ \\
    King&$153.281\pm0.003$&$-1.603\pm0.003$
    & $61.857\pm1.336$& $0.254\pm0.010$ & $R_c: 14\farcm728\pm0\farcm349$, $R_t: 101\farcm578\pm1\farcm832$  \\ 
     & & & & & $R_\mathrm{HL}: 20\farcm788\pm0\farcm295$ \\
    \hline\hline
    \end{tabular}
    \caption{
    The top row, denoted as "Prior", represents a flat prior for each parameter. For instance, $[149.75,156.80]$ (deg.) for R.A. signifies a flat prior within this range. The lower rows provide the structural parameters of Sextans, derived from the likelihood inference of model parameters based on the observed spatial distribution of member stars (refer to Sec.~\ref{sec:structure1d} for further details). With regard to the model profile, we explore the Plummer, exponential, and King models. The model parameters comprise the coordinates of the centre, denoted as RA$_c$ and DEC$c$, the rotational angle ($\theta$), the ellipticity ($\epsilon$), and the characteristic scale radius for each profile: $R_p$ for the Plummer profile, $R_E$ for the exponential profile, and $R_c$ and $R_t$ for the core and tidal radii of the King profile, respectively. Each parameter's central value and the mean $\pm 1\sigma$ error are provided herein. The term $R{\rm HL}$ signifies the half-light radius for each model, which exhibits consistency across all models.
    }
    \label{tab:structural1d}
\end{table*}
\subsection{An estimation of statistical errors in 
Sextans proper motion measurements}
\label{sec:error}
\begin{figure}
    \centering
    \includegraphics[width=0.95\hsize]{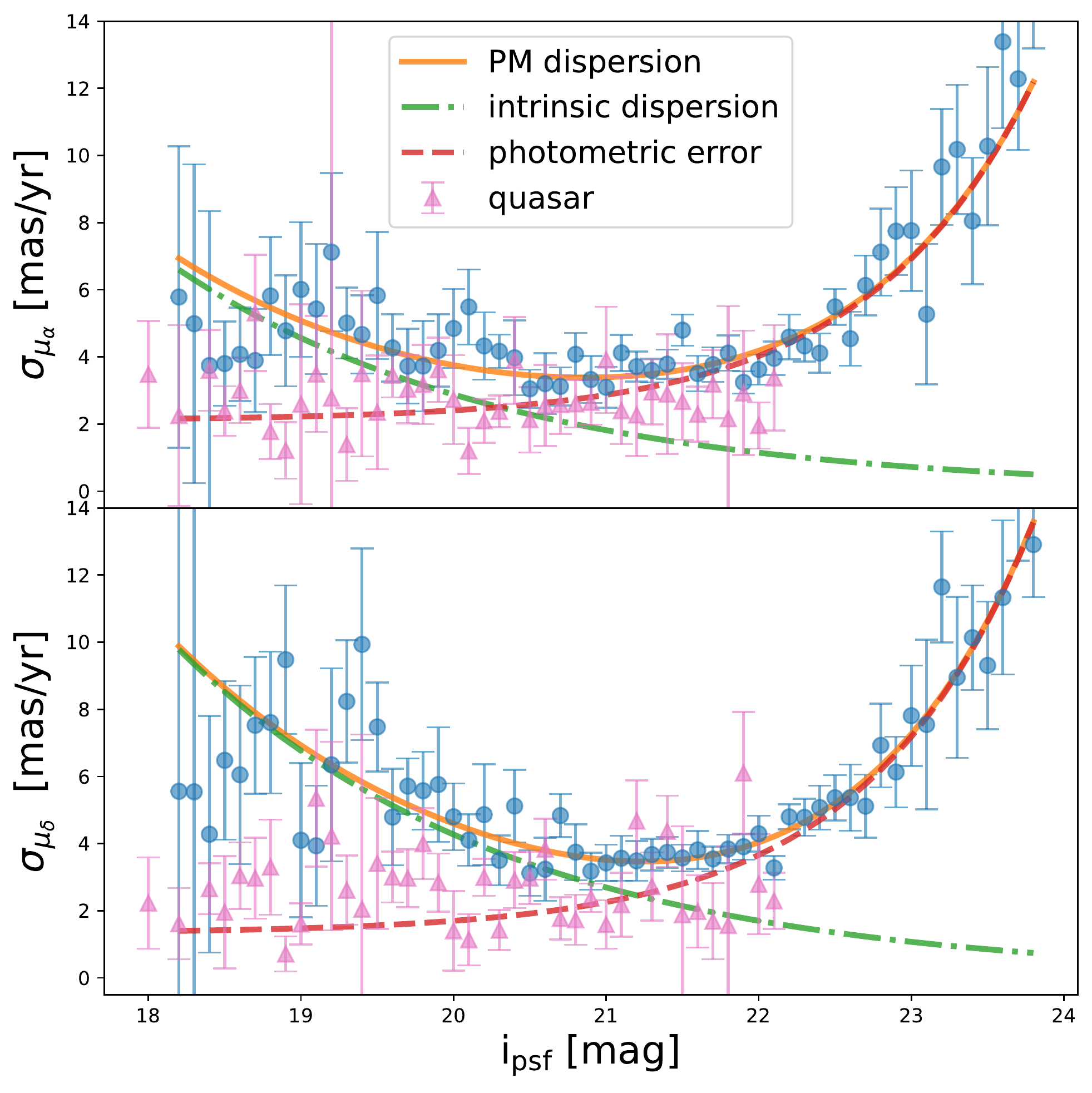}
    \caption{
    The figure presented illustrates the estimation of statistical errors in the proper motion measurements. Circle symbols, accompanied by error bars, denote the dispersion in proper motion in R.A. (upper panel) and Dec. (lower panel) respectively, for stars in each magnitude bin in the field located {\it outside} Sextans, with separation exceeding 1.5deg. (the region beyond the red dashed circle in Fig.\ref{fig:numspa}). The error bars are derived from the standard deviation of 100 bootstrap resampling procedures within each bin. The red dashed line represents a best-fit model premised on the assumption that the dispersion originates from statistical errors in astrometry estimation (the centroid determination) of each star, attributable to photometric errors (refer to the main text for additional details). The green dot-dashed line exhibits a best-fit model presuming the dispersion emerges from the intrinsic, random velocity dispersion of MW halo stars (please consult the main text for further details). The solid orange line sums the contributions from both sources. For comparative purposes, the triangle symbols display the dispersion of apparent proper motions for SDSS quasars in the Sextans field, measured in a manner identical to that applied for the stars.
    }
    \label{fig:errorest}
\end{figure}
Before delving into the results of the Sextans proper motion measurement, we first address the quantification of statistical errors within these measurements. Two primary sources contribute to these statistical errors. The first is a statistical error induced by measurement inaccuracies owing to imperfect astrometry calibration for individual stars. The second source of error originates from the intrinsic random motions of stars, which become particularly significant for foreground MW halo stars. In this section, we elucidate a methodology capable of distinguishing these two sources of error. This method is detailed in Appendix~C of \citet{2021MNRAS.501.5149Q}.

Quasars, being distant point sources, should exhibit no apparent proper motion provided our astrometry is flawless. Therefore, the apparent proper motions of quasars offer an estimate of statistical errors inherent to our proper motion measurements. To this end, we utilise the quasar catalogue from SDSS-DR16Q \citep{2020ApJS..249....3A}. By pairing each of the SDSS quasars with point sources from the HSC-SSP catalogue, we conducted apparent proper motion measurements using the same methodology applied to star proper motion measurements. \re{To elaborate, we adopted the same astrometric recalibration method centred on galaxies as delineated in Sec.~\ref{sec:recalibration}. We consider the quasar catalogue as an independent resource, not utilised within our methodology.}

In Fig.~\ref{fig:errorest}, we demonstrate the variability of proper motions for stars situated beyond a 1.5-degree radius from the Sextans centre. This is contextualised in relation to the luminosity of the stars, where we approximate the dispersion from all celestial bodies positioned within each luminosity bracket. The variability observed can be compared with the dispersion of quasars, analysed in the same manner. It is evident from the figure that the dispersion for quasars is inferior to that for stars within the luminosity bins of $i_{\rm PSF}\lesssim 22$. This implies that there is an extra element contributing to the statistical deviations indicated by the star dispersion. Our analysis is further informed by the red dashed line, representing the best-fit model for the dispersion at $i_{\rm PSF}>22$. This is an expected outcome if we assume that astrometric precision is inversely proportional to flux, as expressed in the equation $1/{\rm flux}\propto 10^{0.4i_{\rm PSF}}$\footnote{For more details, refer to Section~6.2.3 in Robert Lupton's document, wherein he outlines how the statistical precision or error for each star's centroid depends primarily on its flux, or more accurately, on the signal-to-noise ratio of the flux. Given that our proper motion measurements are largely dependent on the precise determination of the object's centroid, the statistical errors of proper motions are contingent upon the accuracy of this centroid determination.} The red dashed curve aptly replicates the measured dispersion at faint magnitudes, in alignment with the statistical measurement errors. The dispersion of quasars aligns with the red dashed curve in brighter magnitude bins, specifically when $i_{\rm PSF}<22$. Conversely, the stars at $i_{\rm PSF}<22$ display larger dispersion than that of quasars or the red-dashed curve.

Consequently, we infer that the dispersion of stars at $i_{\rm PSF}<22$ is influenced by the inherent stochastic proper motions of the Milky Way (MW) halo stars. These halo stars are anticipated to exhibit a constant random velocity dispersion ranging between $80$--100~km~s$^{-1}$ \citep{2010ApJ...716....1B}. If the MW halo stars, chosen based on our colour-magnitude cut (depicted in the right panel of Fig.~\ref{fig:member}), incorporate stars of similar absolute magnitude within each $i_{\rm PSF}$ bracket, then the proper motion dispersion\re{, denoted $\sigma_\mu$, }is predicted to be proportional to the apparent magnitude $i_{\rm PSF}$, as expressed by $\sigma_\mu\propto \sigma_{v}/d\propto 10^{-0.2i_{\rm PSF}}$. The green dot-dashed line in Fig.~\ref{fig:errorest} portrays the best-fit model, where the normalisation parameter of $\sigma_\mu\propto 10^{-0.2i_{\rm PSF}}$ is determined so that the sum of the red dashed and green dot-dashed lines align with the measured dispersion. The solid orange line indicates the best-fit model, which impressively mirrors the calculated dispersion.

Therefore, it is clear that HSC data would allow us to explore the random velocity dispersion of MW halo stars. A more detailed exploration will be featured in a forthcoming study. It is important to highlight that while the spatial variation of solar reflex motions in the Sextans region may contribute to the intrinsic proper motion dispersion, this potential contribution appears negligible within the confined space of our Sextans study area.
\subsection{Measurement results for systemic proper motion of Sextans}
\label{sec:pm_measurements}
We now present the results pertaining to the systemic proper motion measurements of member stars in Sextans. For this purpose, we employ two distinct methods: one that utilises the 1D radial profile of the spatial distribution of member stars, and another that relies on the likelihood analysis of individual stars. \re{It is worth noting that an inherent assumption in our methodology is the absence of spatial variation in the contributions of the proper motions of MW halo stars across the Sextans field.}

To perform the 1D fitting, we commence by measuring the "average" proper motion of stars positioned within each elliptical annulus from the Sextans centre. In doing so, we utilise the elliptical coordinates derived from the best-fit King model as presented in Table~\ref{tab:structural1d}. Following this, we approximate the Sextans' proper motion by minimising the chi-square for each of R.A. or Dec. components:
%
\begin{align}
\chi^2=\sum_{i}\frac{\left[\mu_{\rm meas}(R_i)-\mu_{\rm model}(R_i)\right]^2}{\sigma(R_i)^2},
\label{eq:1dfitting}
\end{align}
where $R_i$ signifies the mean of \re{elliptical} radial separations of stars in the $i$-th radial bin, the summation spans the annulus bins,
$\mu_{\rm meas}(R_i)$ represents the measured average proper motion of stars within the $i$-th bin, and $\mu_{\rm model}(R_i)$ denotes the model proper motion, as defined by:
%
\begin{align}
\mu_{\rm model}(R_i)\equiv 
f(R_i) \mu_{\rm sxt}+\left[1-f(R_i)\right]\mu_{\rm MW}, 
\label{eq:mu_1D_model}
 \end{align}
with 
\begin{align}
f(R_i)\equiv 1-\frac{\Sigma_{b}S(R_i)}{N_{\rm obs}(R_i)}.
\end{align}
Here, $\mu_{\rm sxt}$ denotes the systemic proper motion of Sextans and is a model parameter, $\mu_{\rm MW}$ signifies the systemic proper motion of foreground MW stars, $f(R_i)$ is the fraction of Sextans member stars among the observed stars in the annulus of $R_i$, $\Sigma_b$ represents the number density of foreground MW halo stars, the same quantity used in Eq.~(\ref{eq:likelihood_i}), and $S(R_i)$ is the area of the $i$-th annulus. As we demonstrate later, we adopt radial binning such that each bin contains approximately 300 stars.

For the purposes of this study, we select 7,539 stars situated within the annulus that extends from $R_0$ (the inner radius of the first bin) to $2R_h$ (refer to Fig.~\ref{fig:numspa}) for fitting as our standard choice. The lower limit aids in averting systematic errors as the first annulus is small enough that the galaxy correction is almost entirely accomplished by one grid. The upper limit is defined to incorporate regions where member stars predominate. If we adjust the fitting regions to include stars outside the standard region, the statistical uncertainties of proper motion measurement do not significantly improve, although the central values fluctuate within the statistical errors. In the aforementioned chi-square, we accurately consider the contamination of foreground MW halo stars.

Same in Sec.\ref{sec:structure1d}, we perform MCMC for the estimation. The prior for the proper motions is provided in the "prior" column of Table\ref{tab:propermotion}.
\begin{table}
    \centering
    \begin{tabular}{c|rr|r}\hline\hline
    parameter [mas/yr] & result (1D) & result (2D) & prior \\ \hline
        $\mu_\alpha^\mathrm{sxt}$  & $-0.448\pm0.075$ & $-0.475\pm0.074$ & $[-1, 1]$ \\
        $\mu_\delta^\mathrm{sxt}$  & $0.058\pm0.078$ & $0.008\pm0.078$ & $[-1, 1]$ \\
        $\mu_\alpha^\mathrm{MW}$  & $-2.049\pm0.083$ & -- & $[-10, 0]$ \\
        $\mu_\delta^\mathrm{MW}$  & $-3.507\pm0.103$ & -- & $[-10, 0]$ \\
        \hline\hline
    \end{tabular}
    \caption{The flat prior range of each parameter and the Sextans proper motion measurement results.}
    \label{tab:propermotion}
\end{table}
For $\mu_{\rm MW}$, we approximate the average coherent proper motions of MW halo stars in the regions exterior to Sextans \re{(with an elliptical radius $R_e > 1.5^\circ$)}. We derive $(\mu_{\alpha}^{\rm MW},\mu_{\delta}^{\rm MW})=\left( -2.049, -3.507\right)$[mas~yr$^{-1}$]. We precisely measure the proper motions of MW halo stars, thus we utilise these fixed values in the following fit. We verified that even if we incorporate the impact of the measurement errors of $\mu_{\rm MW}$, the subsequent results remain largely unaltered. To calculate the annulus area $S(R_i)$, we determine the effective area by summing the grids, each with an area of $0.025\times 0.025$~deg$^2$, contained within the annulus, whilst excluding the grids due to bright star masks.

For the statistical uncertainties $\sigma(R_i)$ in the denominator of Eq.~(\ref{eq:1dfitting}), we model it as
\begin{align}
\sigma(R_i)^2=\sigma_{\mu, {\rm meas}}(R_i)^2+\frac{\left[\Sigma_bS(R_i)\right]^2}{N_{\rm obs}^3}
\left(\mu_{\rm sxt}-\mu_{\rm MW}\right)^2,
\end{align}
where $\sigma_{\mu,{\rm meas}}$ is the measurement error in each bin that is estimated from the distribution of proper motions of stars in the $R_i$ bin following the method in \citet{2021MNRAS.501.5149Q} and the second term is from the Poisson error of the number counts of stars in the $R_i$ bin; $\sigma(\delta N_{\rm obs})^2=N_{\rm obs}$. 

\begin{figure}
    \centering
    \includegraphics[width=0.95\hsize]{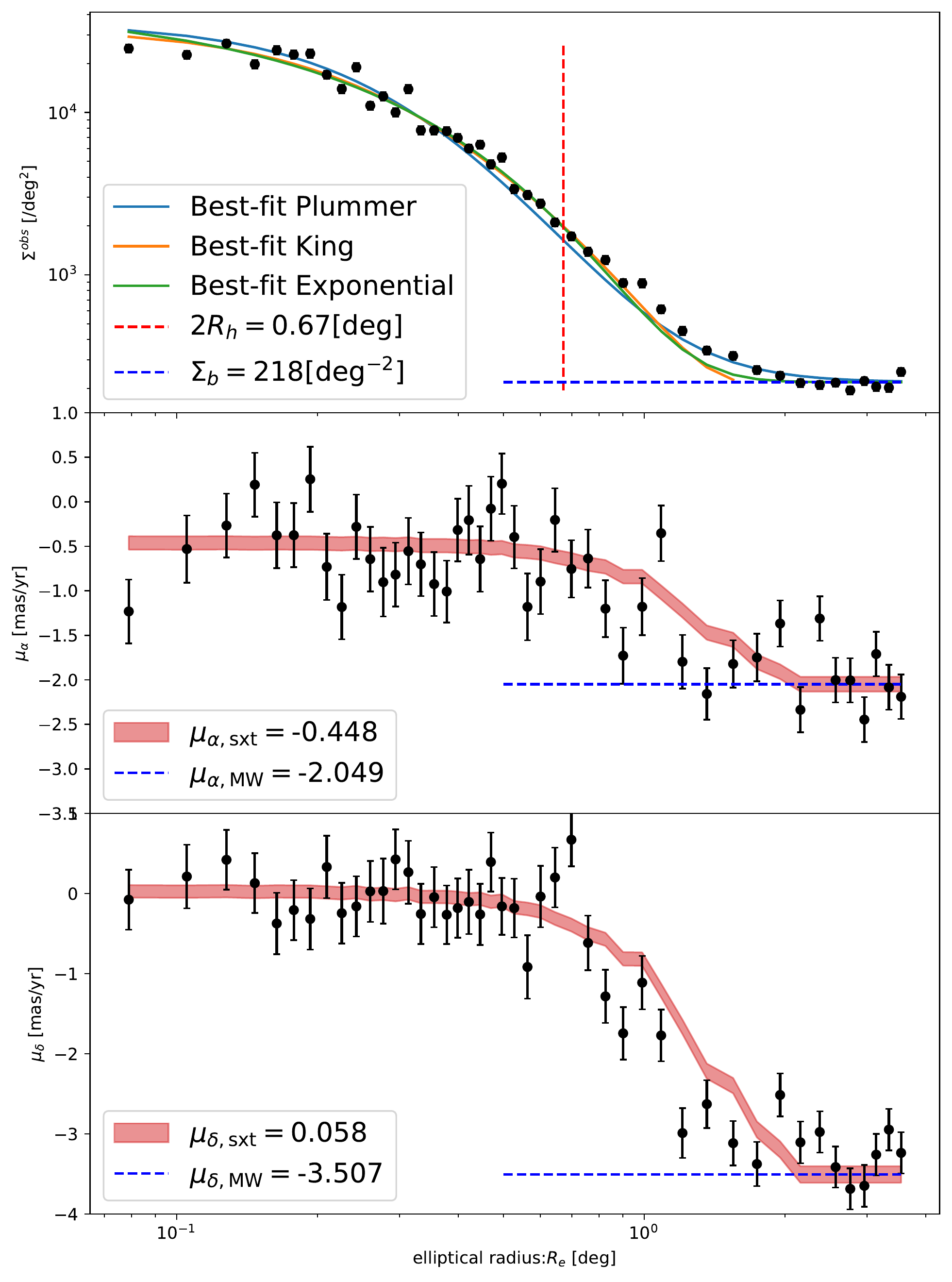}
    \caption{    
    {\it Upper panel}: This panel presents the radial profile of the star density in each elliptical annulus of radius $R_e$, as plotted along the $x$-axis from the Sextans centre. We employ the elliptical coordinates (the axis ratio and position angle) derived from the fitting to the King model detailed in Table~\ref{tab:structural1d}. The solid green, orange and blue lines correspond to the best-fit Exponential, Plummer and King models, respectively. The vertical dashed line represents twice the half-light radius, $2R_h$, within which stars are selected for the proper motion measurements. The horizontal dashed line at outer radii signifies the density of MW halo stars as determined from the field outside Sextans with separations exceeding 1.5~deg. from the Sextans centre. Though the Poisson errors in each bin are shown, the error bars are not visible due to their small size. 
    {\it Middle and lower panels}: The plotted points indicate the calculated systemic proper motion of Sextans, derived from the average of the proper motions of stars in each annulus in both the R.A. and Dec. directions. The errors represent the statistical uncertainties, estimated from the distribution of proper motions of stars in each bin. The red shaded region in each panel signifies the best-fit model for the overall proper motion of Sextans, with the width of the region illustrating the $\pm 1\sigma$ range. The horizontal dashed lines at outer radii represent the coherent proper motion of MW halo stars, as estimated from the average proper motion of stars in the region outside Sextans.
    }
    \label{fig:radpro}
\end{figure}

Fig.~\ref{fig:radpro} illustrates the systemic proper motion measurement of Sextans, derived from the 1D profile fitting (Eq.~\ref{eq:1dfitting}). For large separations where $R_e\gtrsim 2$~deg., the measurements demonstrate a constant, separation-independent proper motion. This suggests that the proper motions are predominantly influenced by a coherent motion of foreground MW halo stars, chiefly due to the apparent reflex motion of stars instigated by the Solar system's proper motion. At smaller radii where $R_e\lesssim 2$~deg., the measurements show scale-dependent profiles of both star distribution and proper motions. The best-fit model prediction and the $1\sigma$ uncertainty are denoted by the red-coloured, shaded lines. These lines effectively replicate the measured proper motions. The resultant proper motion of Sextans is denoted as \val.
As an alternative method, we also perform a maximum likelihood analysis of Sextans proper motion measurement, inspired by  \citet{Pace_2019}. We again use \re{7,539 stars at $R_0 < R_e < 2R_h$ as in the 1D fitting.} We define the likelihood function for the proper motions of all individual stars as 
\begin{align}
{\cal L}^{\rm PM}({\bf d}|
\boldsymbol{\mu}_{\rm model})\equiv \prod_i {\ell}_i^{\rm PM}({\bf d}_i|\boldsymbol{\mu}_{{\rm model}}),
\label{eq:likelihood_2d}
\end{align}
where $\ell_i^{\rm PM}$ is the likelihood for proper motion of the $i$-th star, given by 
\begin{align}
{\ell }^{\rm PM}_i&\equiv f(R_i){\cal N}(\boldsymbol{\mu}_{{\rm meas},i}|\boldsymbol{\mu}_{\rm sxt},\sigma_{\rm sxt})
\nonumber\\
&\hspace{4em}+\left[1-f(R_i)\right]{\cal N}(\boldsymbol{\mu}_{{\rm meas},i}|\boldsymbol{\mu}_{\rm MW},\sigma_{\rm MW}).
\end{align}
Here, the first term represents the likelihood of the star being a member of Sextans, while the second term characterises the likelihood of the star being a foreground MW star. In this context, $\boldsymbol{\mu}{{\rm meas},i}$ refers to the measured proper motion of the $i$-th star, while $\boldsymbol{\mu}{\rm sxt}$ and $\boldsymbol{\mu}{\rm MW}$ correspond to the proper motions of Sextans and MW halo stars, respectively. As in the 1D profile method, we utilise the measured, average proper motions of stars located sufficiently outside of Sextans, while the Sextans proper motions, $\boldsymbol{\mu}{\rm sxt}$, are treated as fitting model parameters. We assume that the proper motion of the $i$-th star follows a Gaussian distribution, denoted as ${\cal N}(\mu_{\rm meas}|\mu_{\rm model}, \sigma_{_{\rm MW}/\rm sxt})$:
%
\begin{align}
{\cal N}(\mu_{\rm meas}|\mu, \sigma_{_{\rm MW}/\rm sxt})\equiv \frac{1}{\sqrt{2\pi}\sigma_{_{\rm MW}/\rm sxt}}
\exp\left[-\frac{(\mu_{\rm meas}-\mu_{\rm model})^2}{2\sigma_{_{\rm MW}/\rm sxt}^2}\right]. 
\end{align}
For the likelihood of a star being a member of Sextans, we employ the dispersion represented by the red dashed curve in Fig.~\ref{fig:errorest}, as a function of $i_{\rm PSF}$, to model the error, denoted $\sigma_{\rm sxt}$. For the likelihood of a star being an MW halo star, we utilise the dispersion symbolised by the orange curve in Fig.~\ref{fig:errorest}, again as a function of $i_{\rm PSF}$, to model the error, denoted $\sigma_{\rm MW}$. To model $f(R_i)$, the probability that a star is a Sextans member, we utilise the measured 1D profile in the upper panel of Fig.~\ref{fig:radpro}. After implementing cubic interpolation on the measured 1D profile, we can estimate $f(R_e)$ at any given separation $R_e$. As a result, the only fitting parameters in Eq.~(\ref{eq:likelihood_2d}) are $\mu_{\rm sxt}^{\alpha}$ and $\mu_{\rm sxt}^{\delta}$, which constitute two model parameters. It's important to note that the likelihoods for each of the R.A. and Dec. components of the Sextans proper motions are separable, enabling us to perform maximum likelihood analysis on the two components separately. The best-fit parameters we derived are \valal. Remarkably, these 2D results align with those derived from the 1D profile fitting, despite the use of identical stars. The 2D likelihood analysis doesn't yield a significantly improved statistical uncertainty in the proper motions compared to the 1D profile method. This suggests that the inclusion of the 2D distribution of proper motions is not significant for the current data and statistical uncertainties.
%
\subsection{Discussion}
\label{sec:discussion}
\begin{figure}
    \centering
    \includegraphics[width=0.95\hsize]{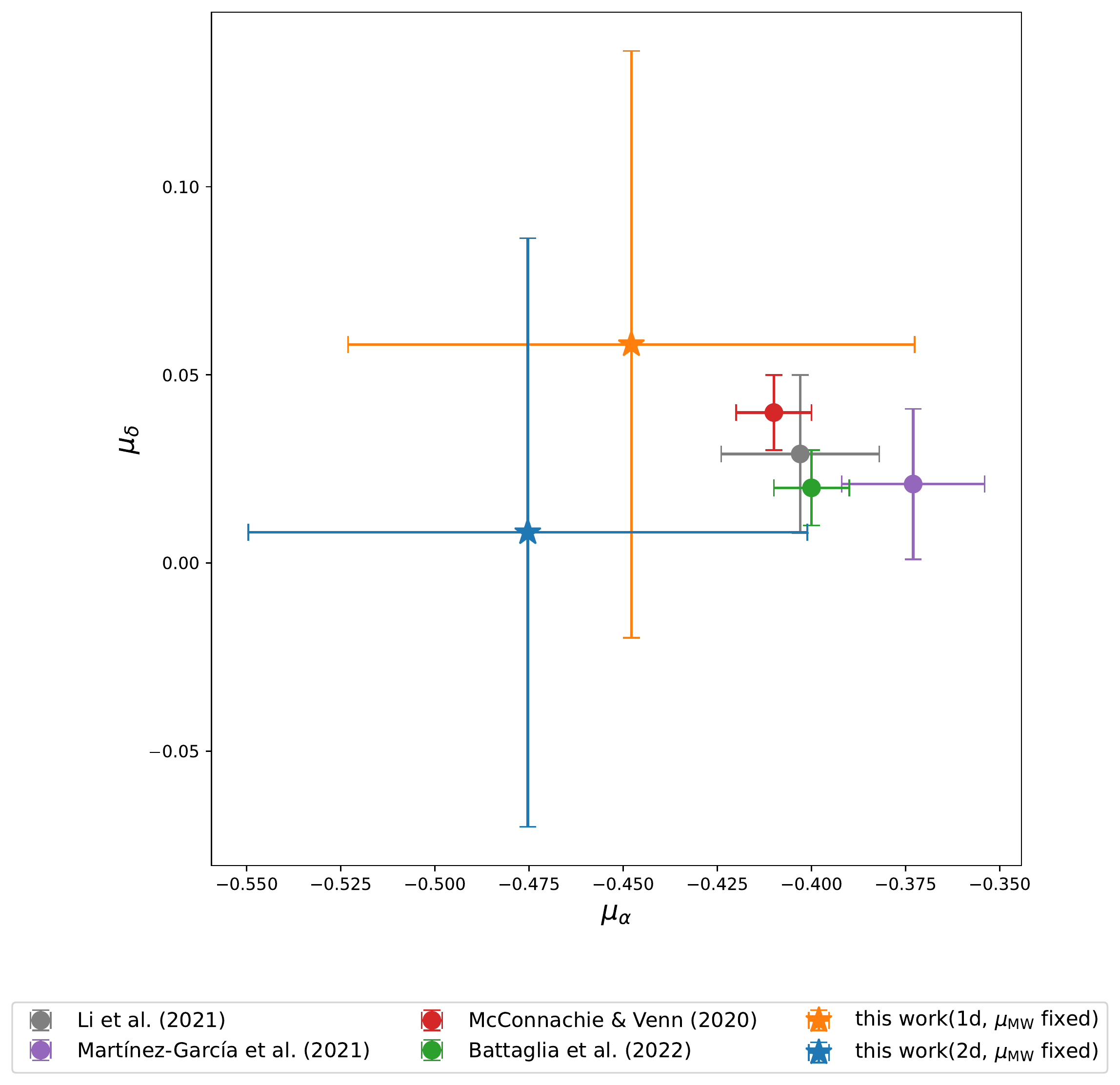}
    \caption{
    This figure provides a comparative view of our proper motion measurements of Sextans (represented by star symbols) in relation to previous works. Our findings are derived from both the 1D profile fitting and the 2D likelihood analysis, as discussed in Sec.~\ref{sec:pm_measurements}. The studies we compare to primarily utilise $Gaia$ data, focusing on fewer, brighter stars in the Sextans region than those investigated in this paper.
    }
    \label{fig:compare}
\end{figure}
In Fig.~\ref{fig:compare}, we compare our measured proper motion with results from previous studies that utilised $Gaia$ EDR3 proper motions of brighter stars in Sextans \citep{McConnachie_2020, Li_2021, 10.1093/mnras/stab1568, 2022A&A...657A..54B}. It is obvious that our conclusions generally align with prior works within the error margins. The $Gaia$ samples are confined to relatively bright BHB and RGB stars with luminosity down to $G \sim 18$, whereas our study extends to faint main sequence stars with brightness as low as $i \sim 24$.

It might seem as though our error margins are considerably larger than those reported by $Gaia$. However, this discrepancy arises from differences in error definition. The error estimates in our study are empirically derived from the dispersion in proper motion, taking into account not only the measurement error but also the intrinsic dispersion in the proper motion of MW foreground stars. Moreover, the error in our work is inversely proportional to the temporal baseline (approximately 2 years). This means that, if we are able to combine future HSC data of Sextans, spanning a longer time baseline of, say, 5 years, with the current data, we could notably decrease the statistical errors.

After adjusting for the influence of Solar reflex motion, we discern that Sextans has a velocity characterised by galactocentric coordinates $\left(V_x, V_y, V_z \right) = \left(-219.6, 82.5, 67.0\right)$~km~s$^{-1}$, equating to a total velocity of approximately $244\ \mathrm{km\ s}^{-1}$. This calculation has been realised by amalgamating our measured proper motion with $\left(D_\mathrm{sxt}, V_\mathrm{r, sxt}\right) = \left(86\ \mathrm{kpc}, 224\ \mathrm{km\ s^{-1}}\right)$ \citep{McConnachie_2020} and making appropriate adjustments for the Solar reflex motion, represented by $\left(V_R, V_\phi, V_Z \right) = \left(-12.9, 245.6, 7.78\right)$~km~s$^{-1}$ \citep{Drimmel_2018}. This velocity value aligns with expectations posited by the MW halo model, assuming a virial mass of the MW to be $M_{\rm vir}\sim 10^{12}$~$M_\odot$ \citep{2018AA...619A.103F}. However, the 3D velocity computed using the Gaia EDR3 proper motion measurement \citep{McConnachie_2020} stands lower, around 230 km s$^{-1}$. Moreover, we discover that Sextans maintains a rather significant peri-centre distance from the galactic centre, estimated at $r_{\rm min}\sim 60$~kpc. This calculation was accomplished by integrating the orbit using {\tt galpy}\footnote{\url{http://github.com/jobovy/galpy}}\citep{Bovy_2015}, a {\tt Python} package specialised in galactic dynamics, along with {\tt McMillan17} \citep{2017MNRAS.465...76M}, one among the potential MW models. This observation suggests that Sextans remains unaffected by tidal disruption, thereby retaining its distinct dark matter structure.
%
\section{Conclusions}
\label{sec:conclusion}
In this study, we have employed two distinct datasets from the Subaru Hyper Suprime-Cam (HSC), the HSC-SSP and the HSC-PI, referred to within this paper in relation to the Sextans field. These datasets were captured at different epochs with a gap of approximately 2.66 years, facilitating the measurement of Sextans' proper motion. The wide field of view coupled with superior image quality (approximately 0\farcs6 seeing) delivered a distinct advantage to our analysis. Primarily, the HSC data afforded us the ability to cleanly distinguish between stars and galaxies. The depth of the data also enabled us to utilise stars down to a magnitude of $i\sim 24$ for these measurements. Subsequently, we used the positions of matched galaxies within the two HSC datasets, captured at different epochs, to recalibrate the astrometry and define the reference frame for our proper motion measurements.

We have meticulously selected probable candidates for Sextans member stars, relying on the colour-magnitude diagram (see Fig.~\ref{fig:member}). Owing to the depth of the HSC, our selection included main-sequence stars; given Sextans' approximate distance of 86 kpc, and a distance modulus of around 19.8. Utilising the spatial distribution of these member candidates, we proceeded to estimate Sextans' structural parameters by fitting the assumed model profile. Our analysis revealed that the distribution of the member stars aligns well with Plummer, exponential, and King profiles. The precision in determining these model parameters was due in large part to the substantial number of member candidates; however, these measured values reveal a minor discrepancy with the results of certain other studies. As these structural parameters reflect the dark matter distribution within Sextans, our findings could hold significance when combined with kinematical structure measurements \citep{2016MNRAS.461.2914H,2020ApJ...904...45H}.

We have measured the proper motion of Sextans and obtained an estimate of \val. While our result aligns fundamentally with previous studies within the margin of error, the precision of our measurement is bound by the time baseline (2.66 years). The key revelation from our research is the efficacy of ground-based data, exemplified by the Subaru HSC, in measuring the proper motion of Sextans at a substantial distance (approximately 86 kpc). This application can be extrapolated to dwarf galaxies at similar distances, provided astrometric solution recalibration is carried out appropriately. The quality of our measurement stands to be improved by incorporating additional HSC data, thereby extending the time baseline. For instance, a twofold increase in the time baseline can enhance the statistical precision of our proper motion measurement by the same factor. This prospect is exciting, particularly with the upcoming LSST survey set to deliver a decade-long observation of the wide-solid angle sky (about 23,000 square degrees). Such comprehensive coverage will encompass many dwarf galaxies, along with other stellar clusters like globular clusters and stellar streams.

Therefore, the measurement of dwarf galaxies' proper motion is a critical metric in testing the assembly history of the Milky Way (MW) halo as well as the character of dark matter structures within these dwarf galaxies and the MW halo. Notably, Sextans serves as a primary dwarf target in upcoming studies with the wide-field-of-view multi-object spectrograph of Subaru, the Prime Focus Spectrograph (PFS), specifically for these objectives \citep{2014PASJ...66R...1T,2020ApJ...904...45H}. Our research provides valuable preliminary information for analyses of dark matter. We are hopeful that our study will provide beneficial guidance for employing ground-based data in future proper motion measurements.

\section*{Acknowledgements}
This work was supported in part by the World Premier International Research centre Initiative (WPI Initiative), MEXT, Japan, JSPS KAKENHI Grant Numbers JP18H05437, JP19H00677, JP20H00172, JP20H00181, JP20H05850, JP20H05855,  JP20H05856, JP21H05448, JP22J11959, JP20H01895, JP21K13909, JP23H04009 and Basic Research Grant (Super AI) of Institute for AI and Beyond of the University of Tokyo. 

This paper is based on data collected at Subaru Telescope and obtained from the SMOKA, which is operated by the Astronomy Data Center, National Astronomical Observatory of Japan.We are honoured and grateful for the opportunity of observing the Universe from Maunakea, which has the cultural, historical, and natural significance in Hawaii.

The Hyper Suprime-Cam (HSC) collaboration includes the astronomical communities of Japan and Taiwan, and Princeton University.  The HSC instrumentation and software were developed by the National Astronomical Observatory of Japan (NAOJ), the Kavli Institute for the Physics and Mathematics of the Universe (Kavli IPMU), the University of Tokyo, the High Energy Accelerator Research Organization (KEK), the Academia Sinica Institute for Astronomy and Astrophysics in Taiwan (ASIAA), and Princeton University.  Funding was contributed by the FIRST program from the Japanese Cabinet Office, the Ministry of Education, Culture, Sports, Science and Technology (MEXT), the Japan Society for the Promotion of Science (JSPS), Japan Science and Technology Agency  (JST), the Toray Science  Foundation, NAOJ, Kavli IPMU, KEK, ASIAA, and Princeton University.

This paper is based [in part] on data collected at the Subaru Telescope and retrieved from the HSC data archive system, which is operated by Subaru Telescope and Astronomy Data Center (ADC) at NAOJ. Data analysis was in part carried out with the cooperation of Center for Computational Astrophysics (CfCA) at NAOJ.  We are honored and grateful for the opportunity of observing the Universe from Maunakea, which has the cultural, historical and natural significance in Hawaii.

This paper makes use of software developed for Vera C. Rubin Observatory. We thank the Rubin Observatory for making their code available as free software at \url{http://pipelines.lsst.io/}. 

The Pan-STARRS1 Surveys (PS1) and the PS1 public science archive have been made possible through contributions by the Institute for Astronomy, the University of Hawaii, the Pan-STARRS Project Office, the Max Planck Society and its participating institutes, the Max Planck Institute for Astronomy, Heidelberg, and the Max Planck Institute for Extraterrestrial Physics, Garching, The Johns Hopkins University, Durham University, the University of Edinburgh, the Queen’s University Belfast, the Harvard-Smithsonian Center for Astrophysics, the Las Cumbres Observatory Global Telescope Network Incorporated, the National Central University of Taiwan, the Space Telescope Science Institute, the National Aeronautics and Space Administration under grant No. NNX08AR22G issued through the Planetary Science Division of the NASA Science Mission Directorate, the National Science Foundation grant No. AST-1238877, the University of Maryland, Eotvos Lorand University (ELTE), the Los Alamos National Laboratory, and the Gordon and Betty Moore Foundation.

This work is based on data collected at Subaru Telescope and obtained from the SMOKA, which is operated by the Astronomy Data Center, National Astronomical Observatory of Japan.

\section*{Data Availability}

The data underlying this article are available in HSC-SSP at \url{https://doi.org/10.1093/pasj/psab122},
SDSS DR16 Quasar at \url{https://doi.org/10.3847/1538-4365/aba623}.
The data were derived from sources in the public domain: 
HSC-SSP at \url{https://hsc-release.mtk.nao.ac.jp/doc/}, 
Subaru Mitaka Okayama Kiso Archive at \url{https://smoka.nao.ac.jp/index.ja.jsp},
SDSS DR16 Quasar at \url{https://www.sdss.org/dr16/algorithms/qso_catalog}.


\bibliographystyle{mnras}
\bibliography{main} 




\appendix
\section{The HSC-SSP sample selection query}
In this appendix, we provide the SQL code used to retrieve the data used in this paper from the HSC-SSP database.
\label{ap:sspsql}
\begin{verbatim}
SELECT
  object_id, 
  f1.i_extendedness_value,
  f2.i_sdsscentroid_ra, f2.i_sdsscentroid_dec,
  f1.g_cmodel_mag, f2.g_psfflux_mag,  
  f1.r_cmodel_mag, f2.r_psfflux_mag, 
  f1.i_cmodel_mag, f2.i_psfflux_mag 
FROM
  s21a_wide.forced AS f1
LEFT JOIN
  s21a_wide.forced2 AS f2 USING (object_id)
WHERE
  f2.i_psfflux_mag<24
  AND f2.i_sdsscentroid_ra > 149.75 
  AND f2.i_sdsscentroid_ra < 156.8
  AND f2.i_sdsscentroid_dec > -2.5 
  AND f2.i_sdsscentroid_dec < 1.7
  AND f1.isprimary
\end{verbatim}

\section{The corner plots of structural property estimations}
\label{ap:wedge}
In this appendix, we provide the corner plots of the MCMC results in different structural profiles: Plummer, Exponential, and King.
\begin{figure}
    \centering
    \includegraphics[width=0.95\hsize]{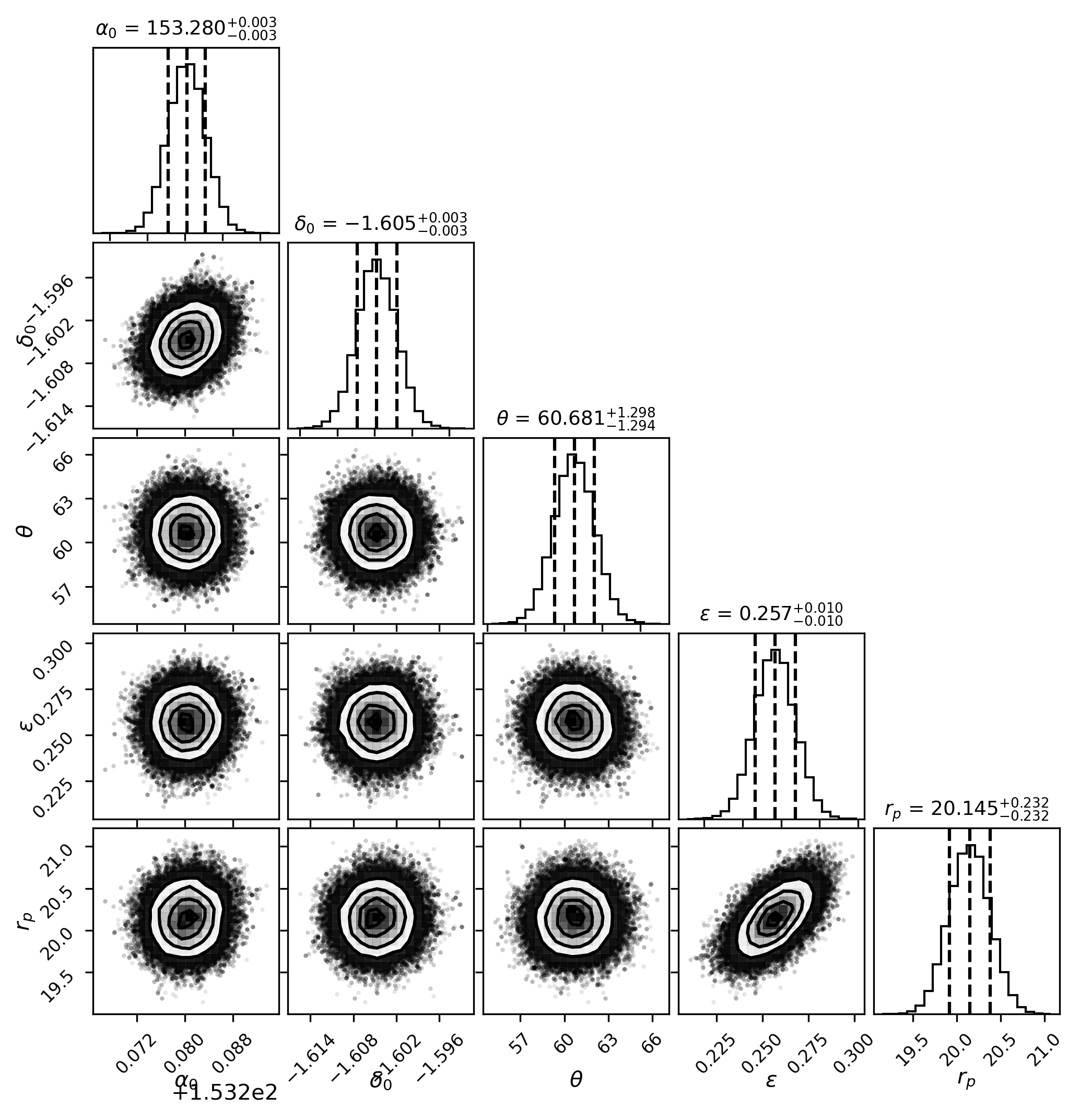}
    \caption{The corner plots of the Plummer structural parameters estimated in Sec.~\ref{sec:structure1d}.
    }
    \label{fig:wedge_plummer}
\end{figure}
\begin{figure}
    \centering
    \includegraphics[width=0.95\hsize]{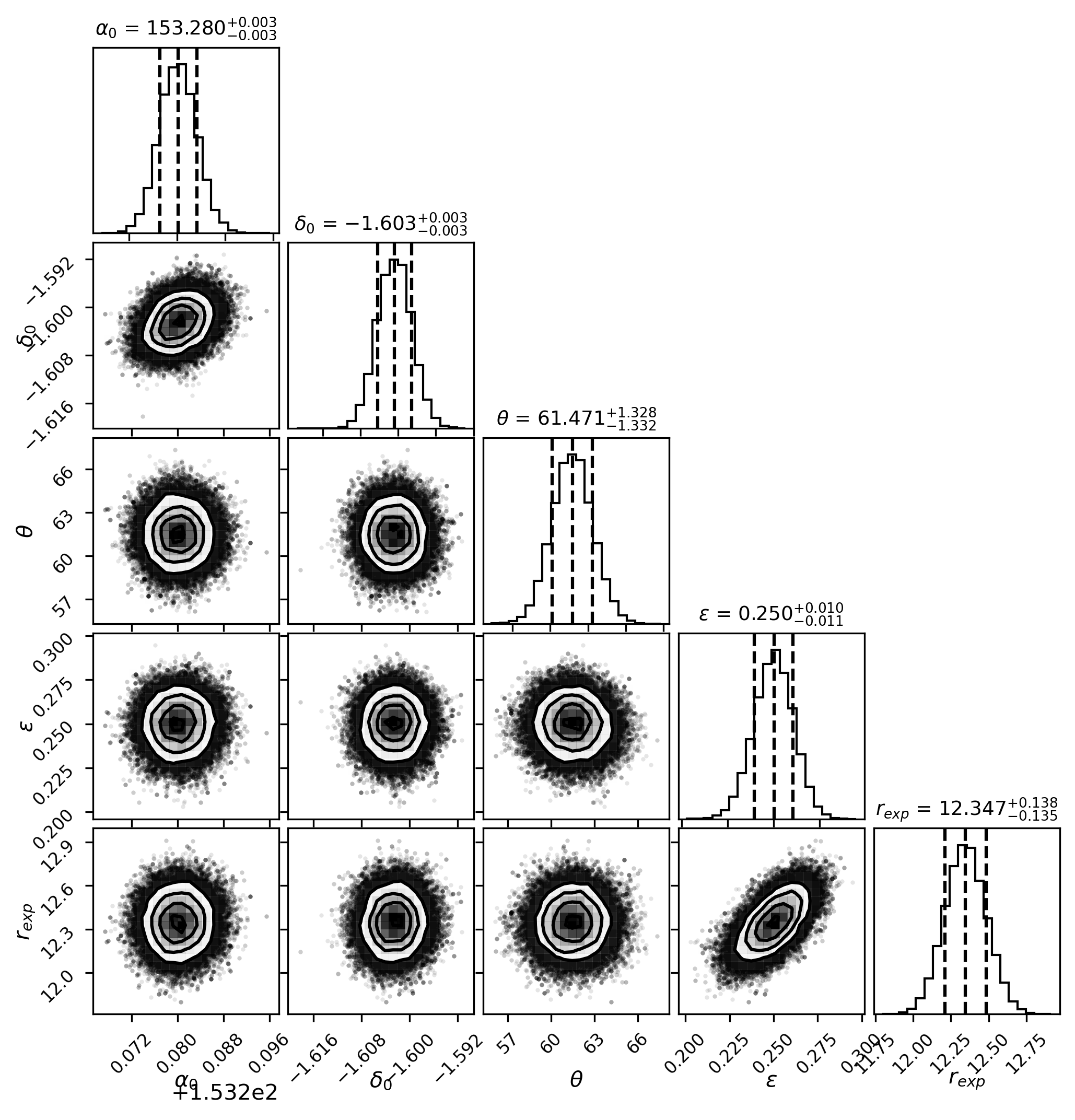}
    \caption{similar for the Exponential model.
    }
    \label{fig:wedge_exp}
\end{figure}
\begin{figure}
    \centering
    \includegraphics[width=0.95\hsize]{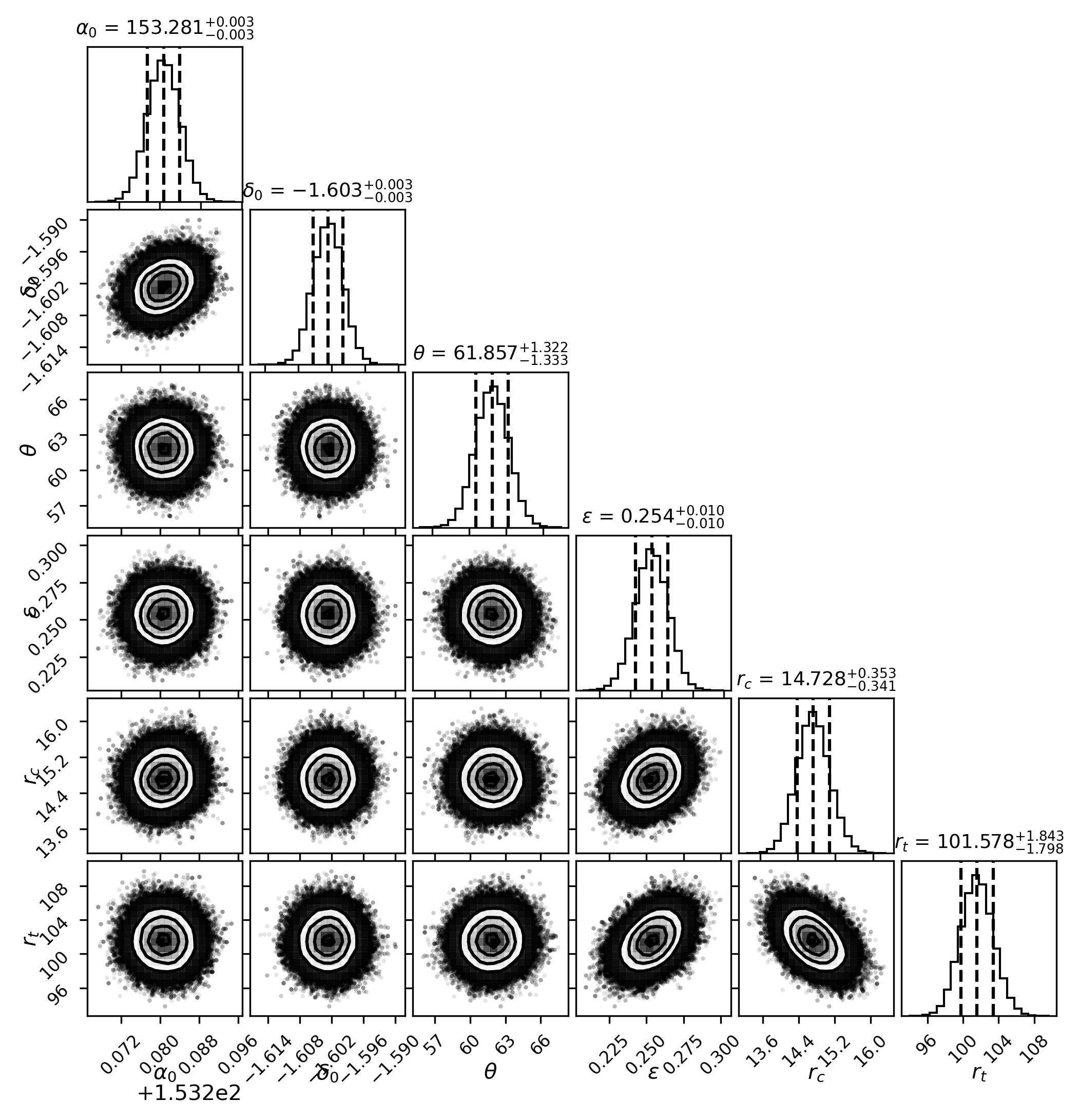}
    \caption{similar for the King model.
    }
    \label{fig:wedge_king}
\end{figure}
%

\bsp	
\label{lastpage}
\end{document}